\RequirePackage{fix-cm}
\documentclass[smallextended]{svjour3}       
\smartqed  
\usepackage{graphicx}
\usepackage[utf8]{inputenc}
\usepackage[english]{babel}
\usepackage[T1]{fontenc}
\usepackage{hyperref}
\usepackage{braket}
\usepackage{amsfonts}
\usepackage{siunitx}
\usepackage{amssymb, amsmath}
\usepackage[subrefformat=parens]{subcaption}

\newcommand{\argmax}{\mathop{\rm arg~max}\limits}

\usepackage{qcircuit}
\usepackage{geometry}
\usepackage{etoolbox}
\usepackage{url}
\apptocmd{\appendices}{\apptocmd{\thesection}{: }{}{}}{}{}
\newcommand{\ggate}[1]{*+<1.4em>{\phantom{#1}} \POS ="i","i"+R; **\dir{-}; \qw}
\usepackage{authblk}

\usepackage[numbers,sort&compress]{natbib}

\usepackage[titletoc,title]{appendix}
\usepackage{etoolbox}
\apptocmd{\appendices}{\apptocmd{\thesection}{: }{}{}}{}{}

\journalname{Quantum Information Processing}

\def\makeheadbox{{%
\hbox to0pt{\vbox{\baselineskip=10dd\hrule\hbox
to\hsize{\vrule\kern3pt\vbox{\kern3pt
\hbox{Quantum Information Processing (2021) 20:293}
\hbox{\href{[Insert doi here]}{https://doi.org/10.1007/s11128-021-03215-9}.}
\kern3pt}\hfil\kern3pt\vrule}\hrule}%
\hss}}}

\begin{document}

\title{Amplitude estimation via maximum likelihood \\on
noisy quantum computer
}

\author{Tomoki~Tanaka \and
        Yohichi~Suzuki \and
        Shumpei~Uno \and
        Rudy~Raymond\and
        Tamiya Onodera\and
        Naoki Yamamoto
        }

\institute{Tomoki~Tanaka \and Yohichi~Suzuki \and Shumpei~Uno \and Rudy~Raymond \and Tamiya~Onodera \and Naoki~Yamamoto
	\at Quantum Computing Center, Keio University, 3-14-1 Hiyoshi, Kohoku-ku, Yokohama, Kanagawa, 223-8522, Japan
	       \and
            Tomoki~Tanaka
             \at Mitsubishi UFJ Financial Group, Inc. and MUFG Bank, Ltd., 2-7-1 Marunouchi, Chiyoda-ku, Tokyo, 100-8388, Japan \\
             Graduate School of Science and Technology, Keio University, 3-14-1 Hiyoshi, Kohoku-ku, Yokohama, Kanagawa, 223- 8522, Japan 
           \and
          Shumpei~Uno
           \at Mizuho Research \& Technologies, Ltd., 2-3 Kanda-Nishikicho, Chiyoda-ku, Tokyo, 101-8443, Japan
           \and
           Rudy~Raymond \and Tamiya~Onodera
            \at IBM Quantum, IBM Research - Tokyo, 19-21 Nihonbashi Hakozaki-cho, Chuo-ku, Tokyo, 103-8510, Japan
            \and
             Naoki~Yamamoto
             \at Department of Applied Physics and Physico-Informatics, Keio University, 3-14-1 Hiyoshi, Kohoku-ku, Yokohama, Kanagawa, 223- 8522, Japan
             \email{yamamoto@appi.keio.ac.jp} 
                 }

\date{Received:21 January 2021 / Accepted: 10 August 2021}

\maketitle

\begin{abstract}
Recently we find several candidates of quantum algorithms that may be implementable in 
near-term devices for estimating the amplitude of a given quantum state, which is a core 
subroutine in various computing tasks such as the Monte Carlo methods.
One of those algorithms is based on the maximum likelihood estimate with parallelized 
quantum circuits. 
In this paper, we extend this method so that it incorporates the realistic noise effect, and 
then give an experimental demonstration on a superconducting IBM Quantum device. 
The maximum likelihood estimator is constructed based on the model assuming the 
depolarization noise. 
We then formulate the problem as a two-parameters estimation problem with respect to 
the target amplitude parameter and the noise parameter. 
In particular we show that there exist anomalous target values, where the Fisher information 
matrix becomes degenerate and consequently the estimation error cannot be improved even 
by increasing the number of amplitude amplifications. 
The experimental demonstration shows that the proposed maximum likelihood estimator 
achieves quantum speedup in the number of queries, though the estimation error saturates 
due to the noise. 
This saturated value of estimation error is consistent to the theory, which implies the validity 
of the depolarization noise model and thereby enables us to predict the basic requirement on 
the hardware components (particularly the gate error) in quantum computers to realize the 
quantum speedup in the amplitude estimation task. 
\keywords{quantum computing \and amplitude estimation \and maximum likelihood estimation  \and depolarizing noise \and IBM Quantum systems}

\end{abstract}

\section{\label{sec:introduction} Introduction}
Ideal and fault-tolerant quantum computers will provide us with game-changing platforms in various area
such as security \cite{shor1999polynomial}, chemical \cite{knill2007optimal, kassal2008polynomial}, and financial engineering \cite{montanaro2015quantum, rebentrost2018quantum, woerner2019quantum, stamatopoulos2019option, martin2019towards, egger2019credit, miyamoto2019reduction}.
Although quantum devices available now are all small and noisy \cite{preskill2018quantum}, the recent progress in hardware \cite{corcoles2019challenges} is certainly going to bridge this gap.
At the same time various software methods, which are expected to lower the hurdle for realizing quantum computing,
are now being developed as well \cite{peruzzo2014variational, farhi2014quantum, temme2017error}.
With careful evaluation of both of these approaches, it is important to have a right prediction
on what applications of quantum computing can be realized and when they might emerge in reality; for instance \cite{jones2012layered, gidney2019factor}.
The quantum volume \cite{cross2019validating} is one reasonable measure along with discussing such predictions.

This paper focuses on the quantum amplitude estimation algorithm \cite{grover1998quantum, Brassard2002},
which can be typically applied to speedup the classical Monte Carlo methods \cite{montanaro2015quantum, rebentrost2018quantum}, etc;
more precisely, in an ideal setup the quantum amplitude estimation algorithm can quadratically reduce the number of samples
and thereby the computation time for Monte Carlo methods.
Because the standard amplitude estimation algorithm is demanding to implement,
due to several technical reasons including the use of many ancilla qubits for quantum Fourier transform and the use of many controlled gate
operations, recently some new techniques that circumvent these challenges have
been developed \cite{suzuki2020, aaronson2019quantum, grinko2019iterative, nakaji2020faster}.
In particular, our approach \cite{suzuki2020} takes the maximum likelihood (ML) method to estimate the amplitude without both the ancilla qubits and the controlled operation conditioned on those qubits,
thereby drastically reducing the number of quantum gates (particularly the controlled NOT gate) involved in this algorithm compared to the standard one;
also the near quadratic speedup was demonstrated in a numerical simulation in the ideal setup.

This paper extends the ML method \cite{suzuki2020} so that it can be used for actual 
noisy quantum computers, and then gives an experimental demonstration on 
a superconducting quantum device. 
The point of the ML method is that, while the true probability distribution generating the data 
is in general unknown, the ML estimator is constructed based on a suitable model distribution. 
In our case, the true distribution is determined from the output state of a noisy 
quantum computer operated under several imperfections that are impossible to perfectly characterize. 
To approximate such unknown noise source for the purpose of constructing the ML estimator, 
in this work, we consider the depolarizing noise model; this is often chosen 
as the minimal or the worst-case noise model 
\cite{koppenhofer2020quantum,magesan2010depolarizing, Noh2020efficientclassical}, 
which is also used for quantifying the gate fidelity of a given quantum computing 
device, i.e., the randomized benchmarking \cite{knill2008randomized, magesan2011scalable, mckay2019three}. 
We also add a fact that the depolarization noise process has a clear mathematical 
merit in that it commutes with an arbitrary unitary operation, which thereby enables 
analytic treatment for our estimation problem. 
We then formulate the problem as a two-parameters estimation problem with 
respect to the target amplitude parameter and the noise parameter. 
This problem formulation introduces a new important aspect to the near-term 
quantum computing field, in the sense that the unavoidable noise coming from 
the realistic imperfection has also to be estimated as a nuisance parameter. 
In our problem, thanks to the property of depolarizing noise, we have an explicit form of 
the Fisher information matrix for discussing the accuracy of estimation and thereby 
derive a formula for specifying the noise level so that near-quadratic speedup is 
achieved to reach a given estimation accuracy. 
Furthermore, the explicit formula of Fisher information matrix reveals the existence 
of anomalous case, where the target parameter cannot be efficiently estimated 
because the Fisher information matrix becomes degenerate. 
Note that, hence, such anomalous target parameter appears only in the multi-parameters 
estimation problem. 
Fortunately we can provide a simple way to circumvent this difficulty.

Below we show the organization of the paper, together with the summary of the results 
obtained in this paper

\begin{description}

\item[Section 2:] 
We take the depolarizing noise model and then formulate the two-parameters estimation problem. 
With this noise model we can have the analytic expression of the Fisher information matrix, which 
gives an asymptotically achievable lower bound of the estimation error. 
This result is further used to derive a condition of the noise level required to have nearly quadratic 
speedup to reach a given estimation error. 
Furthermore, we show the anomalous case where the Fisher information matrix is degenerate 
and consequently the target parameter cannot be efficiently estimated; 
a simple strategy to circumvent this issue is provided.

\item[Section 3:] 
We give an experimental demonstration of our ML method for a simple Monte Carlo 
integration problem, using 2- and 3-qubits IBM Quantum devices \cite{IBMQE, Qiskit}. 
The result is that, for the former case, a quantum speedup in the number of queries 
over the classical one is observed, while the estimation error saturates as the 
query becomes large. 
On the other hand, for the 3-qubit case, we found that the estimation error is 
always bigger than that via the classical method. 
For both cases the saturated value is consistent to the theoretical prediction, implying the validity of the depolarization noise model. 

\item[Section 4:] 
We give the following two discussions. 
First, based on the condition obtained in Section \ref{sec:model} together with the experimental result shown in Section \ref{sec:result}, we discuss the hardware requirements, e.g., 
the error rates of single-qubit and CNOT gates, to achieve a given precision for estimating the value of multidimensional integration. 
Second, the computational complexity of our proposed algorithm is discussed.

\end{description}

\section{Analysis of estimation error under depolarizing noise}
\label{sec:model}

\subsection{Preliminary}
\label{Preliminary}

We briefly review the amplitude estimation algorithm that uses the ML method on parallelized
circuits \cite{suzuki2020}.
This algorithm can estimate the parameter (i.e., the unknown amplitude) quadratically faster
than any classical sampling method to attain a given precision, even without the standard
phase estimation subroutine.
The algorithm mainly consists of two components: one is the amplitude amplification process
\cite{grover1998quantum,Brassard2002}, which is the generalized version of Grover's
quantum search algorithm, and the other is the classical ML estimation part for the data
obtained by the amplitude amplification.
The essential idea of the algorithm is sketched below. \par

The amplitude amplification algorithm initially prepares the state given by
$\ket{\Psi}_{n+1}=\mathcal{A}\ket{0}_{n+1}$, where $\mathcal{A}$ is a unitary operator acting
on the $(n+1)$ qubits.
The operator $\mathcal{A}$ is designed to satisfy
$\mathcal{A}\ket{0}_{n+1} = \sqrt{a}\ket{\tilde{\Psi}_1}_n\ket{1}
	+\sqrt{1-a}\ket{\tilde{\Psi}_0}_n\ket{0}$, where $a\in [0,1]$ is an unknown parameter to be
estimated, and $\ket{\tilde{\Psi}_1}_n$ and $\ket{\tilde{\Psi}_0}_n$ are the $n$-qubit normalized
``good'' and ``bad'' states.
In terms of $\theta_a\in[0,\pi/2]$ satisfying $\sin^2\theta_a=a$, the prepared state
$\ket{\Psi}_{n+1}$ can also be expressed as
\begin{equation}
	\ket{\Psi}_{n+1}
	= \sin{\theta_a}\ket{\tilde{\Psi}_1}_n\ket{1}+\cos{\theta_a}\ket{\tilde{\Psi}_0}_n\ket{0}.
	\label{eq:psi}
\end{equation}
The probability to measure the good state can be amplified by applying the unitary operator
$\mathbf{Q}=-\mathcal{A}\mathbf{S}_0 \mathcal{A}^{-1}\mathbf{S}_{\chi}$ on $\ket{\Psi}_{n+1}$,
where
$\mathbf{S}_{\chi}=\mathbf{I}_{n+1}-2\mathbf{I}_{n}\otimes\ket{1}\bra{1}$
and
$\mathbf{S}_0=\mathbf{I}_{n+1}-2\ket{0}_{n+1}\bra{0}$.
Also $\mathbf{I}_{k}$ is an identity operator on $k$ qubits.
By applying $\mathbf{Q}$ on $\ket{\Psi}_{n+1}$ for $m$ times, we have
\begin{equation}
	\mathbf{Q}^m\ket{\Psi}_{n+1}
	= \sin((2m+1)\theta_a)\ket{\tilde{\Psi}_1}_n\ket{1}
	+ \cos((2 m+1)\theta_a)\ket{\tilde{\Psi}_0}_n\ket{0}.
	\label{eq:Qm}
\end{equation}
This equation tells that the probability for measuring the good state is amplified depending
on the number of repetitions of $\mathbf{Q}$ on $\ket{\Psi}_{n+1}$.
Note that Eq.~\eqref{eq:Qm} is valid only in the absence of noise.

Next, we describe the ML part.
The idea is to estimate the amplitude $a$ using the number of measuring the good state after performing $m_k$ (for $0 \le k \le M$) repetitions of $\mathbf{Q}$.
Now for the ideal state $\mathbf{Q}^{m_k}\ket{\Psi}_{n+1}$, the probability measuring the
good state is $P(m_k ; a)=\sin^2((2m_k +1)\theta_a)$;
then the probability to have the good state $h_k$ times out of the total $N_k$ measurement shots
is proportional to $[P(m_k ; a)]^{h_k} [1-P(m_k ; a)]^{N_k-h_k}$ and accordingly the
likelihood function is
\begin{equation}
	L(\mathbf{h}; a) = \prod_{k=0}^M
	[P(m_k ; a)]^{h_k} [1-P(m_k ; a)]^{N_k-h_k},
\end{equation}
where $\mathbf{h}=(h_0,h_1, \cdots,h_M)$.
The ML estimate of $a$ is then given by
${\rm argmax}_{a} L(\mathbf{h}; a)$, which was proven to achieve the
theoretical lower bound in the estimation error (the detail is described just below).
Note that, of course, the estimated precision depends on the choice of a sequence of integers $\{m_k\}$.
In this paper, we consider the following Linearly Incremental Sequence (LIS) and
Exponential Incremental Sequence (EIS):
\begin{equation}
	\label{LIS and EIS}
	\mbox{(LIS)}~~m_k = k, ~~~~~
	\mbox{(EIS)}~~m_k=\lfloor 2^{k-1}\rfloor, ~\text{for}~0 \le k\le M,
\end{equation}
where in the hereafter we omit the floor notation for simplicity.
To asymptotically achieve a given error $\epsilon$, LIS and EIS need $O(1/\epsilon^{4/3})$
and $O(1/\epsilon)$ total query calls, respectively, while the classical case (i.e., $\forall k: m_k=0$) needs
$O(1/\epsilon^2)$ to achieve the same precision \cite{suzuki2020}.

Lastly we give a general framework of the above-described method for the vector of multiple
parameters $\boldsymbol{\theta}=[\theta_1, \cdots, \theta_K]^\top$ under noisy environment,
which is indeed the scenario studied in this paper ($\bullet^\top$ denotes the transpose).
Let $P(m_k; \boldsymbol{\theta})$ be the probability of measuring the good state for
a density matrix obtained by applying $\mathbf{Q}^{m_k}$ on the state \eqref{eq:psi} under
noisy environment.
Then the likelihood function corresponding to the probability having the good state $h_k$
times in total $N_k$ measurement shots is, similar to the above, given by
\begin{equation}
	L(\mathbf{h};\boldsymbol{\theta})
	=\prod_{k=0}^M \left[P(m_k;\boldsymbol{\theta})\right]^{h_k}
	\left[1-P(m_k;{\boldsymbol{\theta}})\right]^{N_k-h_k},
	\label{likelihood}
\end{equation}
where again $\mathbf{h}=(h_0,h_1, \cdots,h_M)$.
The ML estimate $\hat{\boldsymbol{\theta}}_{\rm ML}$ is defined as
\begin{equation}
	\hat{\boldsymbol{\theta}}_{\rm ML}
	=\argmax_{\boldsymbol{\theta}} L(\mathbf{h};\boldsymbol{\theta})
	=\argmax_{\boldsymbol{\theta}}\ln L(\mathbf{h};\boldsymbol{\theta}).
	\label{maximize_L}
\end{equation}
In general, the estimation error covariance matrix
$\rm{Cov}(\boldsymbol{\hat{\theta}})=\mathbb{E}[(\boldsymbol{\theta}-\boldsymbol{\hat{\theta}})
		(\boldsymbol{\theta}-\boldsymbol{\hat{\theta}})^T ]$, with $\hat{\boldsymbol{\theta}}$
any unbiased estimate, satisfies the Cram\'{e}r--Rao inequality:
\begin{equation}
	{\rm{Cov}}(\boldsymbol{\hat{\theta}})\geq \mathcal{I}^{-1}(\boldsymbol{\theta}),
	\label{CRdef}
\end{equation}
where $\mathcal{I}(\boldsymbol{\theta})$ is the Fisher information matrix defined as
\begin{equation}
	\mathcal{I}(\boldsymbol{\theta})
	=\mathbb{E}\left[
		\left( \frac{\partial}{\partial \boldsymbol{\theta}}\ln L(\mathbf{h};\boldsymbol{\theta}) \right)
		\left( \frac{\partial}{\partial \boldsymbol{\theta}}\ln L(\mathbf{h};\boldsymbol{\theta}) \right)^\top
		\right].
	\label{Fisher info def}
\end{equation}
The expectation value $\mathbb{E} \left[\bullet \right]$ in Eq. (\ref{Fisher info def}) is defined as
\begin{equation}
\mathbb{E}\left[X(\mathbf{h})\right]=\sum_{h_0=0}^{N_0}\sum_{h_1=0}^{N_1}\cdots\sum_{h_M=0}^{N_M}X(\mathbf{h})\left[\prod_{k=0}^{M}
\left(\begin{array}{c}
N_k\\
h_k
\end{array}
\right)P(m_k;\boldsymbol{\theta})^{h_k}(1-P(m_k;\boldsymbol{\theta}))^{N_k-h_k}\right],
\end{equation}
where $X(\mathbf{h})$ is a function of random variables $\mathbf{h}$.
It is known that the ML estimate attains the lower bound of ${\rm{Cov}}(\boldsymbol{\hat{\theta}})$
in Eq.~\eqref{CRdef}, asymptotically in the limit of large samples.
The elements of the Fisher information matrix are given as
\begin{equation}
	\left[\mathcal{I}(\boldsymbol{\theta}) \right] _{i,j}
	=\mathbb{E}\left[
		\left( \frac{\partial}{\partial \theta_i}\ln L(\mathbf{h};\boldsymbol{\theta}) \right)
		\left( \frac{\partial}{\partial \theta_j}\ln L(\mathbf{h};\boldsymbol{\theta}) \right)
		\right].
	\label{Fisher info def comp}
\end{equation}
In the case of the multi-parameter estimation problem, the ML estimation of the $i$-th and $j$-th ($i\neq j$) parameters can be performed independently, if the $(i,j)$ element of the Fisher information matrix is zero. However, otherwise, the $i$-th and $j$-th parameters are correlated, and the estimation of parameter 
of interest may be adversely affected by other nuisance parameters, which is indeed the case in this 
work as shown later.

\subsection{Fisher information in the presence of depolarizing noise}
\label{Section 2.2}

Quantum states are usually disturbed by noise that comes from the system-environment
interaction. 
Then the ideal pure state \eqref{eq:Qm} is replace by a mixed state, which is yet impossible to perfectly identify. 
On the other hand, the ML estimator is based on a model state which may well 
approximate such an unknown true mixed state. 
In this work, we assume the depolarization channel defined by 
\cite{nielsen2002quantum, ji2008parameter}: 
\begin{equation}
	\mathcal{D}(\rho) = p \rho + (1-p) \frac{\mathbf{I}_{n+1}} {d}.
	\label{DC}
\end{equation}
Here $\rho$ is a density matrix, $\mathcal{D}(\rho)$ is a completely positive trace-preserving
(CPTP) map which represents the depolarization of qubits, $1-p$ is the error probability that
qubits are depolarized, and $d$ is the dimension of the quantum system, i.e., $d=2^{n+1}$.
Note that $p$ should also be treated as an unknown parameter, in addition to $a$; hence we 
are studying the two-parameter estimation problem, which is essentially harder compared to 
the one-parameter problem studied in \cite{suzuki2020}.

Now, the CPTP map of the ideal amplitude amplification channel, in terms of the density matrix,
is represented as
\begin{equation}
	\mathcal{Q}(\rho) =\mathbf{Q} \rho \mathbf{Q}^{\dagger}.
	\label{AAC}
\end{equation}
From Eqs. (\ref{DC}) and (\ref{AAC}), the amplitude amplification process in the presence of
noise is thus given by
\begin{equation}
	\rho_{\rm{noise}} = \mathcal{Q}\mathcal{D}  (\rho )=\mathcal{D}\mathcal{Q} (\rho )
	=  p \mathbf{Q} \rho \mathbf{Q}^{\dagger}+ (1-p) \frac{\mathbf{I}_{n+1}} {d}.
	\label{D&AAC}
\end{equation}
Moreover, it is easy to see that $m$ times repetition of this noisy amplitude amplification
process end up with
\begin{equation}
	\rho_{\rm{noise}}^{(m)}=(\mathcal{Q}\mathcal{D} )^{m} (\rho)
	=  p^m \mathbf{Q}^{m} \rho \mathbf{Q}^{\dagger m}+ (1-p^m) \frac{\mathbf{I}_{n+1}}{d}.
	\label{mD&AAC}
\end{equation}
Now the initial state is chosen as $\rho=\ket{\Psi}_{n+1}\bra{\Psi}$, where $\ket{\Psi}_{n+1}$
is given in Eq.~\eqref{eq:psi}.
As in the ideal case, we are interested in the probability of measuring the good state with
which the last qubit is found to be $\ket{1}$, i.e.,
$P(m;\boldsymbol{\theta})={\rm{Tr}}(\rho_{\rm{noise}}^{(m)} E_{1})$, where
$E_1=\mathbf{I}_{n}\otimes\ket{1}\bra{1}$;
also $\boldsymbol{\theta}=[a, \kappa]^\top$ is the vector of unknown parameters,
where $\kappa$ is defined as $\kappa=-\ln p$ which we refer as the noise level of the amplitude amplification process.
By using Eq.~(\ref{eq:Qm}) and ${\rm{Tr}}(\mathbf{I}_{n+1}E_1)=2^n$, this probability is
calculated as
\begin{equation}
	\label{P_D&AAC}
	P(m;\boldsymbol{\theta})
	= P(m; a, \kappa)
	= \frac{1}{2} - \frac{1}{2} \mathrm{e}^{- \kappa m}  \cos(2(2m+1)\theta_a).
\end{equation}
Then the likelihood function \eqref{likelihood} is represented as
\begin{equation}
	\label{Likelihood in Section 3.2}
	L(\mathbf{h}; \boldsymbol{\theta})
	= L(\mathbf{h} ; a, \kappa)
	= \prod_{k=0}^M \left[ P(m_k;a, \kappa)\right]^{h_k}
	\left[1-P(m_k; a, \kappa)\right]^{N_k-h_k}.
\end{equation}

Our task is to estimate $\boldsymbol{\theta}=[a, \kappa]^\top$  via the ML estimate
\eqref{maximize_L}, i.e., $\boldsymbol{\theta}_{\rm ML}
	= {\rm argmax}_{\boldsymbol{\theta}}L(\mathbf{h}; \boldsymbol{\theta})$, 
where $\mathbf{h}=(h_0, h_1, \ldots, h_M)$ is the set of data obtained in the 
experiment. 
As mentioned in the previous subsection, $\boldsymbol{\theta}_{\rm ML}$ 
asymptotically achieves the lower bound in the Cram\'{e}r--Rao inequality 
\eqref{CRdef} 
if the data is generated from the model distribution;  
the Fisher information matrix \eqref{Fisher info def} in this case can now 
be calculated as
\begin{equation}
	\begin{split}
		\mathcal{I}_{1,1}(a,\kappa)
		&= \mathbb{E}\left[
			\left(\frac{\partial}{\partial a}\ln L(\mathbf{h};a,\kappa)\right)^2 \right] \\
		&=\sum_{k=0}^M \frac{N_k(2m_k+1)^2}{\sin^2(2 \theta_a)}
		\frac{4\sin^2\left( 2\left( 2 m_k + 1 \right)\theta_a \right)}
		{e^{2\kappa m_k}-\cos^2\left(2\left( 2 m_k+1 \right)\theta_a\right)}, \\
		\mathcal{I}_{1,2}(a,\kappa) = \mathcal{I}_{2,1}(a,\kappa)
		&= \mathbb{E}\left[
			\left(\frac{\partial}{\partial a}    \ln L(\mathbf{h};a,\kappa)\right)
			\left(\frac{\partial}{\partial\kappa}\ln L(\mathbf{h};a,\kappa)\right)\right] \\
		&=\sum_{k=0}^M \frac{N_km_k(2m_k+1)}{\sin(2\theta_a)}
		\frac{\sin\left( 4\left( 2 m_k + 1 \right)\theta_a \right)}
		{e^{2\kappa m_k}-\cos^2\left(2\left( 2 m_k+1 \right)\theta_a\right)}, \\
		\mathcal{I}_{2,2}(a,\kappa)
		&= \mathbb{E}\left[
			\left(\frac{\partial}{\partial\kappa}\ln L(\mathbf{h};a,\kappa)\right)^2 \right] \\
		&=\sum_{k=0}^M N_km_k^2
		\frac{\cos^2\left( 2\left( 2 m_k + 1 \right)\theta_a \right)}
		{e^{2\kappa m_k}-\cos^2\left(2\left( 2 m_k+1 \right)\theta_a\right)}.
	\end{split}
\end{equation}
Recall that $a$ is the parameter of our true interest;
from Eq.~\eqref{CRdef}, the estimation error of $a$ is lower bounded by the $(1,1)$
element of the inverse of the above Fisher information matrix as
\begin{equation}
	\label{2dim and 1dim CR inequality}
	\epsilon = \sqrt{\mathbb{E}\Big[ (a - \hat{a} )^2 \Big]}
	\ge \sqrt{\left(\mathcal{I}(a,\kappa)^{-1} \right)_{1,1}}
	=: \epsilon_{\rm min}(a,\kappa).
\end{equation}
We use $\epsilon_{\rm min}(a,\kappa)$ to discuss the condition on the noise 
level $\kappa$ to satisfy a required estimation precision. 
These topics are studied in detail in the next subsection.

Lastly we remark on other noise models. 
The most general (Markovian) noise model, with multiple parameters, can be 
represented by the Kraus superoperator. 
However, in general this does not commute with Grover operator, and as a result 
we cannot obtain an analytic expression of Cram\'{e}r--Rao lower bound. 
To discuss the estimation accuracy in such a case, several numerical methods have 
been developed in the field of quantum metrology 
\cite{ji2008parameter,escher2011general,yuan2017quantum}. 
Note that even for a special type of noise channel other than depolarization, it 
is still difficult to have an analytic expression of Cram\'{e}r--Rao lower bound. 
For instance, in Ref.~\cite{brown2020quantum} considering the amplitude damping and 
dephasing noise model in the same amplitude estimation problem yet with known 
noise strength, the estimation error was evaluated numerically.

\subsection{Achievable estimation error and required depolarizing noise level}
\label{sec:2.3}

\begin{figure}
	\centering
	\includegraphics[width=11.5cm]{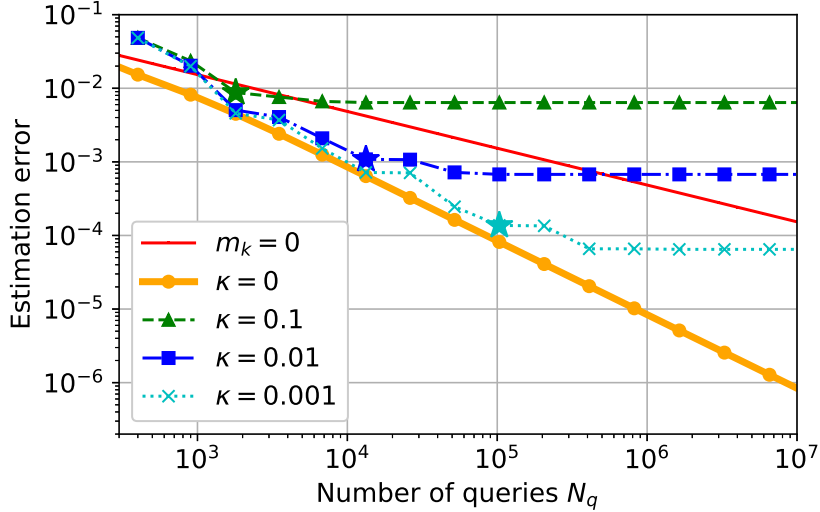}
	\caption{
	Cram\'{e}r--Rao lower bound $\epsilon_{\rm min}(a,\kappa)$ on the estimation error
	of $a$, versus the total number of queries.
	The red line corresponds to the classical case $\forall k: m_k=0$, while
	the other lines are the quantum cases of EIS with several values of noise level $\kappa$.
	See the first paragraph of Section~\ref{sec:2.3} for the details.
	}
	\label{queryVSerror}
\end{figure}

Figure~\ref{queryVSerror} shows the Cram\'{e}r--Rao lower bound of the estimation error,
$\epsilon_{\rm min}(a,\kappa)$, versus the total number of query calls,
$N_{\rm{q}}=\sum_{k=0}^{M}N_k(2 m_k+1)$.
In particular in the classical case, it is given by
$N_{\rm{q}}=\sum_{k=0}^{M}N_k$.
The target value is chosen to be $a=\sin^2\theta_a=0.375$, and $N_k=100$ for all $k$.
The (red) thin solid line represents the lower bound in the classical case.
The other lines represent the lower bounds obtained when using the amplitude amplification
with the EIS for several noise level $\kappa$;
the (yellow) thick solid line is the bound without noise ($\kappa=0$);
the (green) dashed line with triangles, the (blue) dash-dotted line with squares, and the
(light blue) dotted line with crosses are the lower bound under depolarizing noise
$\kappa=10^{-1}, 10^{-2}$, and $10^{-3}$, respectively.
Recall that, in the ideal case $\kappa=0$, the number of query calls needed to reach
a specified value of $\epsilon$ is $N_{\rm q}\sim O(1/\epsilon)$, i.e., the Heisenberg-scaling.
We note that a similar dependence of $\epsilon_{\rm min}(a,\kappa)$ on the magnitude
of $\kappa$ is observed for many cases of $a$, but there exist cases such that
$\epsilon_{\rm min}(a, \kappa)$ takes much bigger values than those shown in
Fig.~\ref{queryVSerror} even when $\kappa$ is sufficiently small;
see Section~\ref{sec:anomaloustarget} about this special case.

A notable point observed in Fig.~\ref{queryVSerror} is that, even under the depolarizing noise model, the
estimation error $\epsilon_{\rm min}(a,\kappa)$ decreases in nearly the Heisenberg-scaling law
up to $N_{\rm q}\sim 10^4$ and $N_{\rm q}\sim 10^5$ for the cases $\kappa=0.01$ and
$\kappa=0.001$, respectively.
However, the error does not decrease anymore, even by using more queries.
In other words, $\epsilon_{\rm min}(a,\kappa)$ gets saturated at those points of $N_{\rm q}$,
and thus the algorithm has to be stopped\footnote{
This saturation always occurs if $\kappa\neq 0$, which can be proven using the d'Alembert's
ratio test together with the inequality
$\epsilon_{\rm min}(a,\kappa) \geq \left(
	\mathcal{I}_{1,1}(a,\kappa)\right)^{-1/2}
	\geq \left( \sum_{k=0}^{M}
	\frac{N_k 4 (2 m_k +1)^2}{\sin^2{2\theta_a}} \frac{e^{-2 \kappa m_k }}{1-e^{-2 \kappa m_k }}\right)^{-1/2}$.
}.
The maximum number of queries within which the Heisenberg-scaling is guaranteed can be formally characterized as follows.
That is, even under depolarizing noise model with $\kappa$, the Heisenberg-scaling is nearly
preserved if the number of operations of $\mathcal{Q}$, $m_k$, is smaller than $\bar{m}$ defined as the maximum integer satisfying the following inequality \cite{ji2008parameter}:
\begin{equation}
	\label{mlim}
	(2\bar{m} + 1) (1 - \mathrm{e}^{-\kappa} ) \leq 1.
\end{equation}
This condition is derived as a sufficient condition to guarantee that the probability to measure the good state is not affected by the noise in the limit of large samples.
The star marks in Fig.~\ref{queryVSerror} are the total query calls $\bar{N}_{\rm q}$ corresponding to $\bar{m}$ for given $\kappa$, showing that in fact the estimation error does not obey the
Heisenberg-scaling law even by calling more queries
than $\bar{N}_{\rm q}$.
Note that $\bar{m}$ given in Eq.~\eqref{mlim} was originally derived in the one-parameter
setting (that is, the case where $\kappa$ is known), while $\epsilon_{\rm min}(a,\kappa)$
is a function of the two-parameter Fisher information matrix; despite of this gap $\bar{m}$
certainly captures the point of maximum number of query calls up to which the estimation
error decreases according to the Heisenberg-scaling.

\begin{figure}
	\centering
	\includegraphics[width=12.0cm]{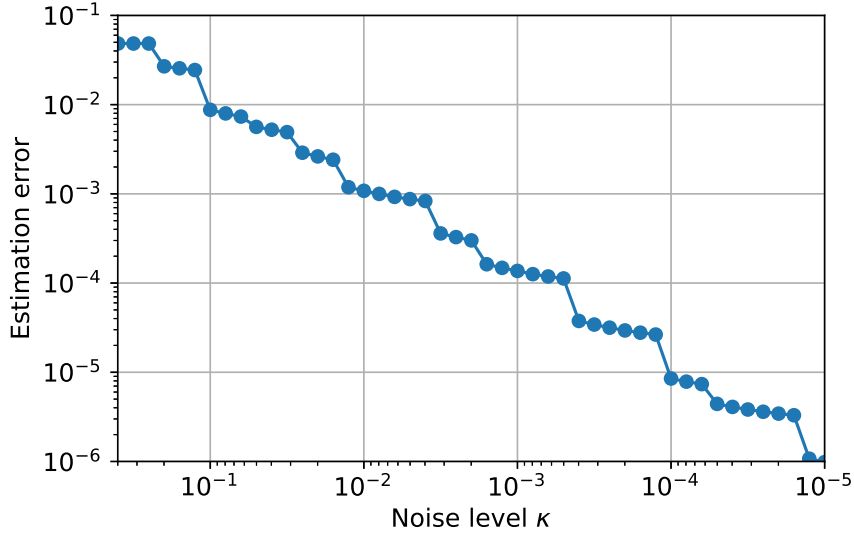}
	\caption{
		Relationship between the noise level $\kappa$ and the achievable error
		$\epsilon$, in the case $a=0.375$.
		The ML algorithm is with $\forall k: N_k=100$ and EIS.}
	\label{KVSerror}
\end{figure}

Based on the above-described fact, we obtain the condition on the noise level $\kappa$
so that the ML algorithm reaches a specified estimation error $\epsilon$ with the number
of query calls of the order $O(1/\epsilon)$, i.e., the Heisenberg-scaling, even under
the noisy environment.
Fig.~\ref{KVSerror} yields such condition; this is the relation between $\kappa$ and
the error at $\bar{N}_{\rm q}$, for the case $a=\sin^2\theta_a=0.375$, $\forall k: N_k=100$,
and EIS.
For instance, if we want to reach the estimation error $\epsilon=10^{-4}$ using
$O(1/\epsilon)$ query calls, then we need $\kappa$ to be smaller than $\sim 10^{-3}$.
Importantly, Fig.~\ref{KVSerror} indicates the ``quasi linear relation" between a specified
$\epsilon$ and the required value of $\kappa$; that is, if we need to decrease $\epsilon$ to
$\epsilon/10$ then $\kappa$ should be improved to simply $\kappa/10$.
Note that this quasi linear relation is expected to hold from Eq.~\eqref{mlim}, which leads to $2\bar{m}+1=1/\kappa$ when
$\kappa$ is small, together with the Heisenberg-scaling
$\bar{N}_{\rm q} = O(1/\epsilon)$, although Eq.~\eqref{mlim} was proven only in the one-parameter setting.
The condition on $\kappa$ can be further converted to that on the gate fidelity of
elementary gates constructing the quantum circuit.
That is, Fig.~\ref{KVSerror} represents the minimum hardware specification required to
apply the quantum amplitude estimation method to solve a concrete problem such as a
Monte Carlo integration task demonstrated later; a more detailed discussion on the
hardware requirement will be given in Section~\ref{sec:evaluation}.

The above discussion as well as Figs.~\ref{queryVSerror} and \ref{KVSerror}
surely depend completely on the mathematical model described in Section~\ref{Section 2.2}.
The real quantum system must not perfectly coincide with this model. The resulting ML
estimate is then not guaranteed to reach the Cram\'{e}r--Rao lower bound discussed here;
also the quasi linear relation observed in Fig.~\ref{KVSerror} could be changed.
That is, only with the materials posed up to now, we still could not say that Fig.~\ref{KVSerror}
serves as a guide for discussing the condition on $\kappa$ to reach a specified estimation
error with Heisenberg-scaling law in the real world.
This fully motivates us to execute some detailed experiment to see if the theory described
above is consistent to the experimental result and thereby verify its usability 
to the real quantum computing applications; 
this topic will be discussed in Section \ref{sec:result}.

Note that there is still a room for improvement on update strategy of $m_k$. For instance, if we employ $m_k=\lfloor r^{k-1} \rfloor$ where $r$ is a real number which satisfies $r>1$, the value of $\bar{N}_{\rm{q}}$ changes depending on the value of $r$. This indicates that the achievable estimation error in the presence of noise can be reduced by changing the update strategy. 
More importantly, especially for the two-parameters problem considered in 
this paper, the estimation precision can be severely limited depending on the 
choice of $r$, which will be discussed in the next subsection.

\subsection{Anomalous target value that induces a large
	estimation error}
\label{sec:anomaloustarget}

\begin{figure}
	\centering
	\includegraphics[width=11.0cm]{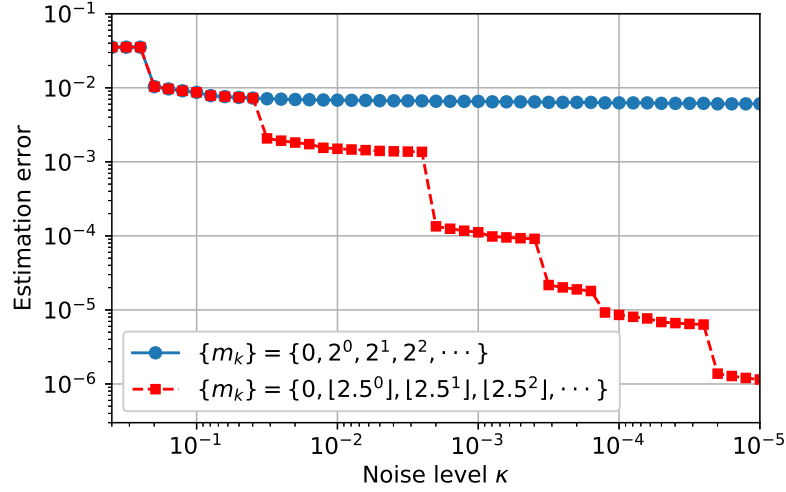}
	\caption{
		Lower bound of the estimation error
		$\epsilon_{\rm min}=\sqrt{(\mathcal{I}^{-1})_{1,1}}$ versus  the noise level $\kappa$.
		The target value is chosen as $a=\sin^2(\pi/8)$. The solid (blue) and dotted (red) lines are obtained when using $\{m_k\}=\{0,2^0,2^1,2^2,\cdots\}$ and $\{m_k\}=\{0,\lfloor{2.5^0}\rfloor,\lfloor{2.5^1}\rfloor,\lfloor{2.5^2}\rfloor,\cdots\}$, respectively.
		$\lfloor\cdot\rfloor$ is the floor function.
	}
	\label{fig:queryVSerror_anomalous_target}
\end{figure}

The previous discussion 
is based on the
following two conditions; that is, the quasi linear relation between $\kappa$ and $\epsilon_{\rm{min}}(a,\kappa)$, and the Heisenberg scaling of the total number of queries $N_{\rm{q}}=\sum_{k=0}^M N_k(2 m_k+1)$ with respect to a given estimation error $\epsilon$ in
the range $m_k\lesssim \bar{m}$.
However, these conditions does not always hold.
For instance, when the target value $a$ is $a=\sin^2 (\pi/8)$,
the estimation error $\epsilon_{\rm{min}}(a,\kappa)$ does not
obey the quasi linear relation with respect to $\kappa$, as shown by the blue solid line in Fig.~\ref{fig:queryVSerror_anomalous_target}.
That is, even when the noise $\kappa$ is sufficiently small, a precise
estimate of $a$ is not possible.
This is due to the multi-parameter estimation setting, where
in general the estimation error covariance matrix of the parameters
$\boldsymbol{\theta}$ satisfies the Cram\'{e}r--Rao inequality
\eqref{CRdef}, i.e.,
${\rm{Cov}}(\boldsymbol{\hat{\theta}})\geq \mathcal{I}^{-1}(\boldsymbol{\theta})$, where
$\mathcal{I}(\boldsymbol{\theta})$ is the Fisher information
matrix.
Clearly, if ${\rm det}\, \mathcal{I}(\boldsymbol{\theta})$ is nearly
zero at a certain point of $\boldsymbol{\theta}$, then the
estimation of those parameters has to be inefficient.
Actually in the above-described example, our Fisher information
matrix $\mathcal{I}(a, \kappa)$ is nearly degenerate at around
$a=\sin^2 (\pi/8)$; this is the reason why the quasi linear
decrease of $\epsilon$ with respect to $\kappa$ does not hold
in this case.
In this section, we investigate this ``anomalous target" point
of $a$ in detail.
But before moving forward, we would like to emphasize that
this analysis never happen in the 1-parameter setting.
That is, as stated in Section 1, to achieve quantum advantage on available
noisy quantum devices, the noise parameter has to
be incorporated into the parameters to be estimated and accordingly
such a singular point analysis needs to be carried out.

\begin{figure}
	\centering
	\includegraphics[width=14.0cm]{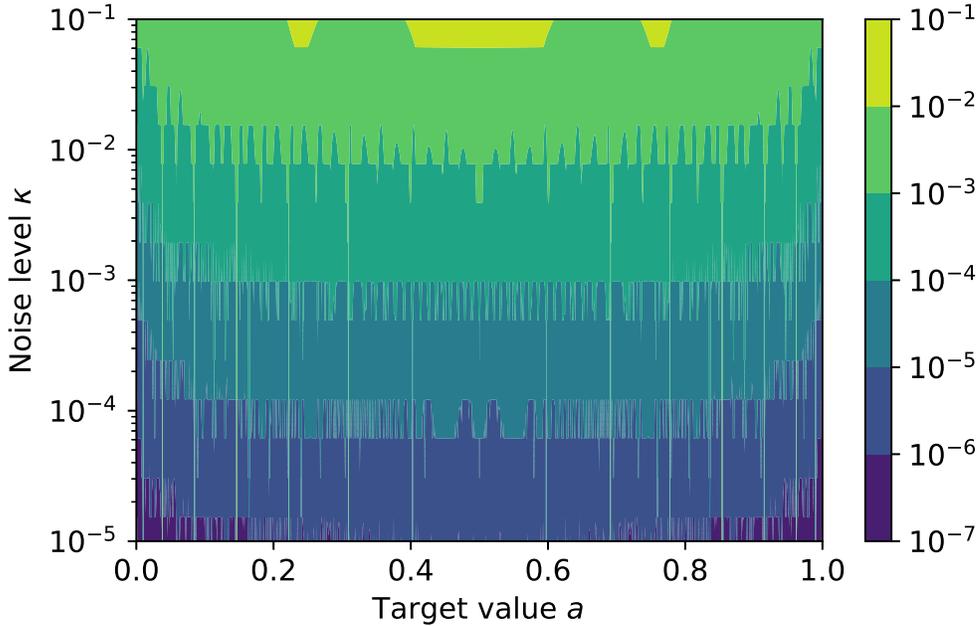}
	\caption{
		Contour plot of the lower bound of the estimation error,  $\epsilon_{\rm min}=\sqrt{(\mathcal{I}^{-1})_{1,1}}$, as
		a function of the target value $a$ and the noise level $\kappa$.
	}
	\label{fig:Countour2DimFisher}
\end{figure}

First, to see how likely such an anomalous target of $a$ exists, we
plot the lower bound of the estimation error
$\epsilon_{\rm min}=\sqrt{(\mathcal{I}^{-1})_{1,1}}$, as a function
of $a$ and $\kappa$ in Fig.~\ref{fig:Countour2DimFisher}.
This contour plot shows that, for almost all target values, the
estimation error $\epsilon_{\rm min}$ almost linearly decreases
with respect to $\kappa$;
that is, $\epsilon_{\rm min}$ decreases approximately by one order
of magnitude, as $\kappa$ decreases by one order of magnitude.
However, there exist anomalous target values of which estimation
errors are insensitive to the value of $\kappa$;
for instance, at around $a=\sin^2(\pi/8)=0.146$, we observe a
long spike where $\epsilon_{\rm min}$ takes almost the same value
in the range $[1\times 10^{-2}, 1\times 10^{-3}]$ regardless of $\kappa$.

\begin{figure}
	\begin{minipage}{\hsize}
		\begin{center}
			\includegraphics[width=12.0cm]{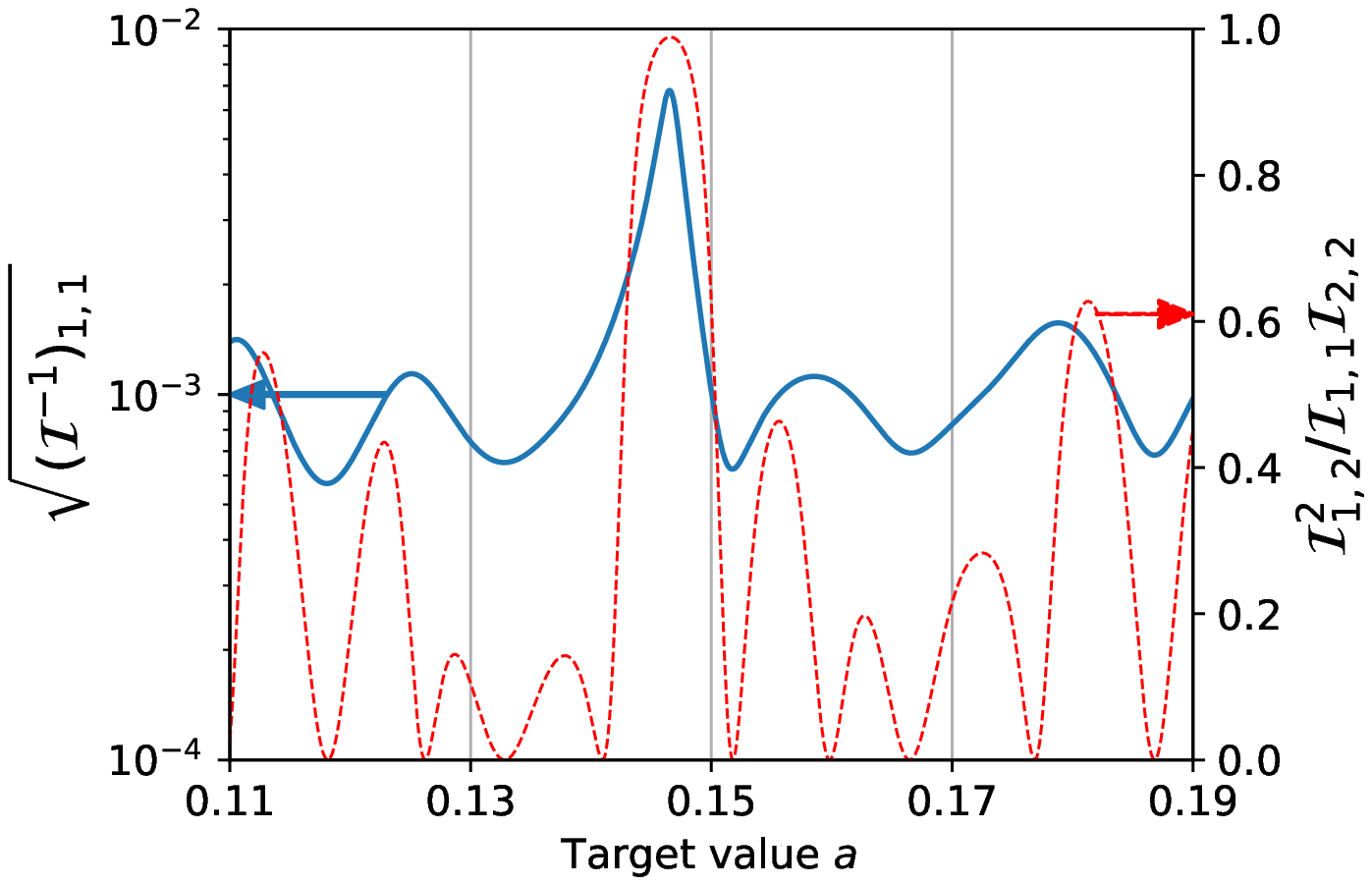}\\(A) $\kappa=10^{-2}$
		\end{center}
	\end{minipage}
	\begin{minipage}{\hsize}
		\begin{center}
			\includegraphics[width=12.0cm]{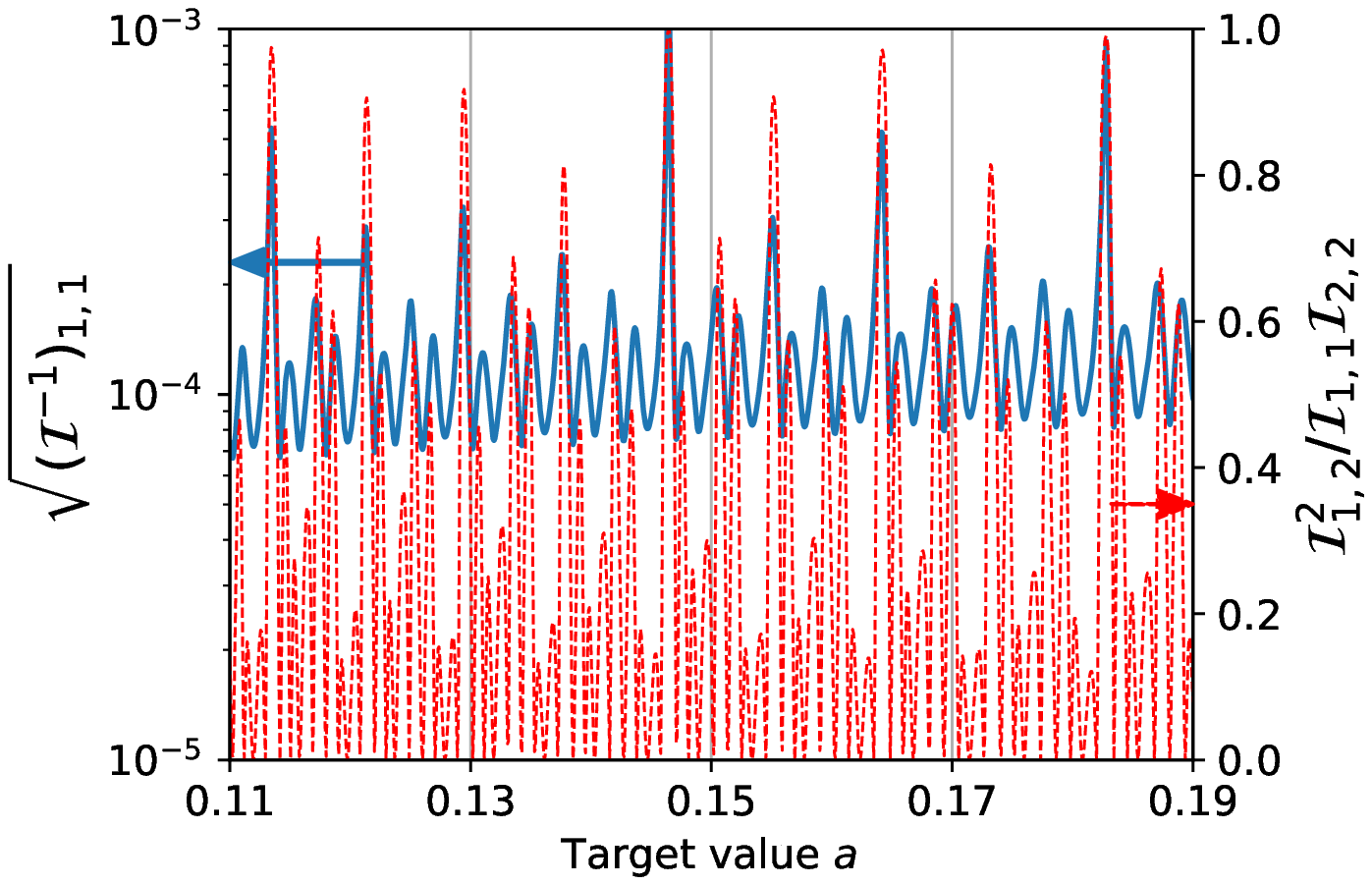}\\{(B) $\kappa=10^{-3}$}
		\end{center}
	\end{minipage}
	\caption{
	Estimation error
	$\epsilon_{\rm min}=\sqrt{(\mathcal{I}^{-1})_{1,1}}$
	represented with the solid (blue) line and the anomality
	$\beta = \mathcal{I}_{1,2}^2/\mathcal{I}_{1,1}\mathcal{I}_{2,2}$
	represented with the dotted (red) line,
	as a function of the target value $a$.
	The noise parameter $\kappa$ is fixed to
	(A) $\kappa=10^{-2}$ and (B) $\kappa=10^{-3}$.
	}
	\label{fig:DeterminantVS2DimFisher}
\end{figure}

This insensitivity of $\epsilon_{\rm min}$ at a certain point of
$a$ is originated from the fact that, as mentioned in the first
paragraph of this section, the Fisher information matrix $\mathcal{I}(a, \kappa)$ is nearly degenerate at around $a$.
To closely look into the relation between the estimation error
and the degeneracy of $\mathcal{I}(a, \kappa)$, in
Fig.~\ref{fig:DeterminantVS2DimFisher} we plot $\epsilon_{\rm min}$
with the solid blue line and the ``anomality"
$\beta = \mathcal{I}_{1,2}^2/\mathcal{I}_{1,1}\mathcal{I}_{2,2}$
with the dotted red line, as a function of the target value $a$.
The noise parameter $\kappa$ is fixed to
(A) $\kappa=10^{-2}$ and (B) $\kappa=10^{-3}$.
Note that $0<\beta\leq 1$ and $\beta=1$ is equivalent to
${\rm det}\, \mathcal{I}(a, \kappa)=0$; hence, $\beta \sim 1$ means
that such $(a, \kappa)$ are difficult to estimate precisely.
Both of (A) and (B) of Fig.~\ref{fig:DeterminantVS2DimFisher}
indeed show that, at around the anomalous target of $a$ where
$\beta=1$, the estimation error takes a relatively large value.

The existence ratio of anomalous target values can be quantitatively
analyzed in terms of the {\it linear density} defined as follows.
Here, $N=10^5$ samples of $a$ are randomly chosen from the uniform distribution on $[0,1]$, and and the ratio of $a$ satisfying
$\beta > 0.9$ is computed; the linear density is given by this ratio.
Table~\ref{tab:RatioOfAnomalousKappa} shows the linear density
for several values of $\kappa$.
Importantly, the linear density is almost independent of the
value of $\kappa$; it is about $1\% \sim 2\%$ regardless of $\kappa$.
This is due to the composite of the following two properties of the  anomalous targets: (i) the number of anomalous segments satisfying $\beta > 0.9$ is inversely proportional to $\kappa$, and (ii) the length of each anomalous segment is proportional to $\kappa$. Since the linear density of anomalous targets is approximately the product of the number of the anomalous regions with the average length of anomalous regions, it is almost insensitive to $\kappa$.

It should be noted that the linear density of anomalous targets takes finite value in the limit of $\kappa\to 0$, while the anomalous targets themselves are not present in the case of $\kappa=0$. This is essentially originated from whether one has the information of the noise or not, i.e. the Fisher information $\mathcal{I}_{1,1}$ can be employed to obtain the lower bound of the estimation error in the absence of the noise, however $(\mathcal{I}^{-1})_{1,1}$ should be used if the noise level is unknown.

\begin{table}[htb]
	\caption{
		The linear density of anomalous targets in the range $[0,1]$.
		The statistical error of the linear density is calculated under
		the assumption that the linear density obeys a binomial distribution.}
	\label{tab:RatioOfAnomalousKappa}
	\begin{center}
		\begin{tabular}{cc}
			\hline
			$\kappa$   & the linear density of anomalous targets $a$         \\
			\hline
			$10^{-10}$ & 1.56\hspace{1mm}$\pm$\hspace{1mm}0.04\hspace{1mm}\% \\
			$10^{-9} $ & 1.84\hspace{1mm}$\pm$\hspace{1mm}0.04\hspace{1mm}\% \\
			$10^{-8} $ & 1.90\hspace{1mm}$\pm$\hspace{1mm}0.04\hspace{1mm}\% \\
			$10^{-7} $ & 1.56\hspace{1mm}$\pm$\hspace{1mm}0.04\hspace{1mm}\% \\
			$10^{-6} $ & 2.01\hspace{1mm}$\pm$\hspace{1mm}0.04\hspace{1mm}\% \\
			$10^{-5} $ & 1.70\hspace{1mm}$\pm$\hspace{1mm}0.04\hspace{1mm}\% \\
			$10^{-4} $ & 1.67\hspace{1mm}$\pm$\hspace{1mm}0.04\hspace{1mm}\% \\
			$10^{-3} $ & 2.02\hspace{1mm}$\pm$\hspace{1mm}0.04\hspace{1mm}\% \\
			$10^{-2} $ & 1.28\hspace{1mm}$\pm$\hspace{1mm}0.04\hspace{1mm}\% \\
			$10^{-1} $ & 0.00\hspace{1mm}$\pm$\hspace{1mm}0.00\hspace{1mm}\% \\
			\hline
		\end{tabular}
	\end{center}
\end{table}

Finally, we propose two approaches to avoid the anomalous case.
The first one is based on the observation that the determinant of the Fisher information matrix  changes depending on the choice of the sequence $\{m_k\}$ of the
amplitude amplification.
Therefore, if the underlying target value is detected to be anomalous,
then we can try another sequence $\{m_k\}$ to avoid the
anomaly.
For instance, when  $\{m_k\} = \{0,\lfloor{2.5^0}\rfloor,\lfloor{2.5^1}\rfloor,\lfloor{2.5^{2}}\rfloor,\cdots\}$, the quasi linear relation between $\kappa$ and $\epsilon_{\rm{min}}$ is recovered even when $a=\sin^2(\pi/8)$, as
shown with the dotted red line in Fig.~\ref{fig:queryVSerror_anomalous_target}.
This is a clear evidence showing that by suitably choosing $\{m_k\}$ the determinant of the Fisher information matrix does not get smaller.
Our view is that it might be possible to detect the anomalous target
by calculating the empirical Fisher information, which eventually
allows us to tune the sequence and thereby avoid the anomality.

The second approach is by modifying the amplitude to be estimated.
After detecting the anomalous target, we introduce an extra
ancilla qubit; then $R_y(\phi)$ rotation (i.e., the single qubit
rotation around the y-axis with a fixed parameter $\phi$) is applied
to the ancilla qubit as follows:
\begin{equation}
	\begin{split}
		\mathcal{A}\ket{0}_{n+2} =& \sqrt{a}\ket{\tilde{\Psi}_1}_n\ket{1}\ket{0}
		+\sqrt{1-a}\ket{\tilde{\Psi}_0}_n\ket{0}\ket{0} \\
		\xrightarrow{Ry(\phi)}&
		\cos(\phi)\sqrt{a}\ket{\tilde{\Psi}_1}_n\ket{1}\ket{0}
		+\sin(\phi)\sqrt{a}\ket{\tilde{\Psi}_1}_n\ket{1}\ket{1} \\
		&+\cos(\phi)\sqrt{1-a}\ket{\tilde{\Psi}_0}_n\ket{0}\ket{0}
		+\sin(\phi)\sqrt{1-a}\ket{\tilde{\Psi}_0}_n\ket{0}\ket{1}.
	\end{split}.
\end{equation}
By estimating the probability that the last two-qubit state is $\ket{1}\ket{1}$, we could avoid the anomalous target problem,
because the amplitude to be estimated is modified from $a$ to $a\sin^2(\phi)$.

\section{Experiment with a real quantum computing device \label{sec:result} }

This section is devoted to show experimental result using the real-backend 
device of IBM Quantum Systems called \textit{ibmq\_valencia} \cite{IBMQE}, 
to evaluate the estimation performance of the proposed ML estimate and thereby 
the validity of the employed depolarization model.
In particular we consider the Monte Carlo integration problem, whose computational
(sampling) cost can be quadratically reduced via the amplitude estimation algorithm
\cite{montanaro2015quantum, rebentrost2018quantum, woerner2019quantum, stamatopoulos2019option, martin2019towards, egger2019credit, miyamoto2019reduction}.
In this section, we begin with a brief explanation on the target integration problem, followed
by showing the execution results of the real device for two-qubit and three-qubit cases.

\subsection{Monte Carlo integration via amplitude estimation}
\label{sec:simpleexample}

Let us consider a general 1-dimensional integration $\mathbb{E}[f(x)] = \int_{0}^{1}dx q(x)f(x)$,
where $f:[0,1]\to[0,1]$ is any real-valued smooth function and $q:[0,1]\to[0,1]$ is the probability
distribution function which satisfies $\int_0^1dx q(x) = 1$.
This quantity can be in practice obtained by calculating the approximation
\begin{equation}
	\label{eq:DefinitionS}
	S(f) = \sum_{j=0}^{2^n-1}p(x_j)f(x_j),
\end{equation}
where $x_j$ is defined as $x_j = (j + 1/2)/2^n$ and $p$ is the probability mass function defined as $p(x_j) = \int_{x_j-1/2^{n+1}}^{x_j+1/2^{n+1}} q(x) dx $.
It should be noted that there is an approximation error due to the discretization of $f(x)$, i.e., $S(f)\neq\mathbb{E}[f(x)]$. In our analysis, however, we evaluate the error between
$S(f)$ and the value obtained by Monte Carlo integration in order to assess the performance of our algorithm on a real quantum device. The amplitude estimation algorithm
is run via the following operators:
\begin{align}
	\mathcal{P}\ket{0}_n = \sum_{j=0}^{2^n-1}\sqrt{p(x_j)}\ket{j}_n, ~~
	\mathcal{R}\ket{j}_n\ket{0} = \ket{j}_n\left(\sqrt{f(x_j)}\ket{1}+\sqrt{1-f(x_j)}\ket{0}\right).\label{eq:OperatorIntegration}
\end{align}
By applying these operators to the $(n+1)$-qubit initial state, $\Ket{0}_n\Ket{0}$, we obtain
\begin{equation}
	\label{RP0 state}
	\begin{split}
		\mathcal{R}(\mathcal{P} \otimes \mathbf{I}_{1})\Ket{0}_n\Ket{0}
		&=\sum_{j=0}^{2^n-1}\sqrt{p(x_j)}\ket{j}_n\left(\sqrt{f(x_j)}\ket{1}+\sqrt{1-f(x_j)}\ket{0}\right)
		\\
		&=\sqrt{S(f)} \ket{\tilde{\Psi}_1}\ket{1}+\sqrt{1-S(f)} \ket{\tilde{\Psi}_0}\ket{0},
	\end{split}
\end{equation}
where $\ket{\tilde{\Psi}_1}$ and $\ket{\tilde{\Psi}_0}$ are defined as
\begin{align}
	\ket{\tilde{\Psi}_1}
	=\frac{1}{\sqrt{S(f)}}\sum_{j=0}^{2^n-1}\sqrt{p(x_j) f(x_j)}\ket{j}_n, ~~
	\ket{\tilde{\Psi}_0}
	=\frac{1}{\sqrt{1-S(f)}}\sum_{j=0}^{2^n-1}\sqrt{p(x_j)(1-f(x_j))}\ket{j}_n.
\end{align}
This is exactly the state of the form \eqref{eq:psi}.
Thus, the ideal amplitude estimation algorithm gives an approximation of $S(f)$ with the
precision $\epsilon$, with only $O(1/\epsilon)$ queries.

In this paper, we consider the simple integration $ \int_0^{1} \sin^2(bx) \ dx$ with
$b$ a constant, which is approximated as
\begin{equation}
	S(f) = \sum_{j=0}^{2^n-1} \frac{1}{2^n} \sin^2\left(b x_j\right). \label{eq:SumSimpleSin}
\end{equation}
In this case the operators $\mathcal{P}$ and $\mathcal{R}$ in
Eq.~\eqref{eq:OperatorIntegration} can be implemented only with Hadamard gates and
controlled $Y$-rotation gates, as shown in Fig.~\ref{2qubitsC}.

\subsection{Experimental result for the two-qubits case \label{sec:2result} }

We now show the experimental result of applying the ML algorithm in the real device, to
compute Eq.~\eqref{eq:SumSimpleSin}.
In this subsection, we consider the simple case $n=1$, meaning that the integration is
approximated via the discrete sum $S(f)$ having only two domain values $x=0$ or $x=1$,
in which case Eq.~\eqref{eq:SumSimpleSin} takes $b=\pi/20$, that is, the value $S(f)=a=\sin^2\theta_a=0.0077$.
Also this setting means that we need only two qubits; in the experiment we chose the 0-th
and 1-st qubit of \textit{ibmq\_valencia}.

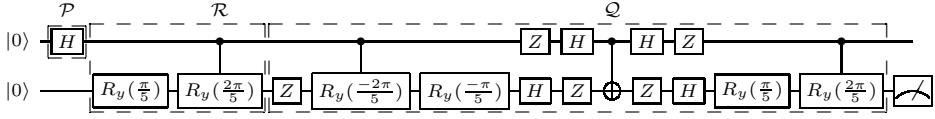
\begin{figure}[tb]
	\centering
	\leavevmode
	\scriptsize
	\Qcircuit @C=.5em @R=.8em {
	&\mathcal{P} & & \mathcal{R} & &  & & & & \mathcal{Q}\\
	\lstick{ \ket{0} } & \gate{H} &  \ggate{} & \ctrl{1} & \ggate{}  & \ctrl{1} & \qw & \gate{Z} & \gate{H}  & \ctrl{1} & \gate{H} & \gate{Z} & \qw & \ctrl{1} & \qw \\
	\lstick{ \ket{0} } & \ggate{} &\gate{R_y(\frac{\pi}{5})} & \gate{R_y(\frac{2\pi}{5})} & \gate{Z} & \gate{R_y(\frac{-2\pi}{5})} & \gate{R_y(\frac{-\pi}{5})} &  \gate{H} & \gate{Z} & \targ & \gate{Z} & \gate{H} & \gate{R_y(\frac{\pi}{5})} & \gate{R_y(\frac{2\pi}{5})} & \meter
	\gategroup{2}{2}{2}{2} {.3em}{--}
	\gategroup{2}{3}{3}{4} {.3em}{--}
	\gategroup{2}{5}{3}{14} {.3em}{--}
	}
	\caption{
		The 2-qubit circuit of the unitary operators $\mathcal{Q}$ and
		$\mathcal{R}(\mathcal{P} \otimes I)$, for computing the probability
		$P(m;\boldsymbol{\theta})$ where the target value $S(f)$ is given in
		Eq.~\eqref{eq:SumSimpleSin} with $n=1$, and $b=2\pi/5$.}
	\label{2qubitsC}
\end{figure}

First, we show the experimental result of the quantum algorithm to
compute the probability $P(m_k;a, \kappa)$ given in Eq.~\eqref{P_D&AAC}; recall that
\begin{equation}
	\label{prob in Section 3.2}
	\begin{split}
	P(m_k;a, \kappa) = {\rm{Tr}}(E_{1} \rho_{\rm{noise}}^{(m_k)} )
	&= {\rm{Tr}}\Big[ (\mathbf{I}_1\otimes\ket{1}\bra{1})
		(\mathcal{Q}\mathcal{D} )^{m_k} (\ket{\Psi}_2\bra{\Psi}) \Big] \\
	&= \frac{1}{2} - \frac{1}{2} \mathrm{e}^{- \kappa m_k}  \cos(2(2m_k+1)\theta_a),
	\end{split}
\end{equation}
and this is used to construct the likelihood function
\eqref{Likelihood in Section 3.2}.

\begin{figure}[tb]
	\centering
	\includegraphics[width=12.0cm]{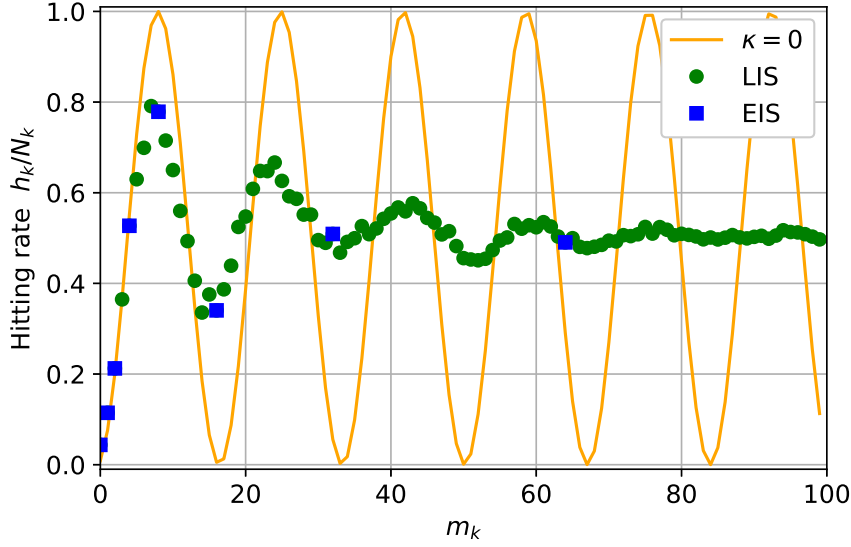}
	\caption{
		The hitting rate of ancilla qubit being $1$ (i.e., $h_k/N_k$), versus the number
		of $\mathcal{Q}$ in amplitude amplifications, $m_k$.
		The (green) round and (blue) square points show the experimental result obtained by
		running the ML algorithm with LIS and EIS, respectively.
		The (orange) line is the analytic result of the ideal probability without noise.
		To mark each point, we performed $N_k = 8,192$ measurements for all $k$.  }
	\label{LIS}
\end{figure}

The quantum circuit to execute the unitary operators $\mathcal{Q}$ and
$\mathcal{R}(\mathcal{P} \otimes I)$ is shown in Fig.~\ref{2qubitsC}.
Here $\ket{\Psi}_2=\mathcal{R}(\mathcal{P} \otimes I)\Ket{0}\Ket{0}$ is given by Eq.~\eqref{RP0 state}.
In Fig \ref{LIS}, the green round points are plotted by computing the hitting rate of ancilla qubit being $1$ (i.e., the hitting rate of measuring the good state, which corresponds to $P(m_k; a, \kappa)$),
for the LIS setting.
Note that these points cover the points of the EIS setting which are marked with the blue square points depicted in Fig.~\ref{LIS}. Importantly, the figures show that the hitting rate has a trend of exponentially-decaying 
oscillation and approaches to $0.5$ as the number of $\mathcal{Q}$ in amplitude amplification, 
$m_k$, increases. 
As a minimal model, we take the depolarizing noise \eqref{DC} to model this decayed oscillation; 
actually the resulting probability distribution \eqref{prob in Section 3.2} well interpolates all the 
points obtained in the experiment, as shown in Fig.~\ref{LIS}. 
Also, for reference, the analytic result of the ideal probability in the absence of noise,
i.e., the case $\kappa=0$, is depicted with the orange line in the Fig.~\ref{LIS}.

\begin{figure}[tb]
	\centering
	\includegraphics[width=12.0cm]{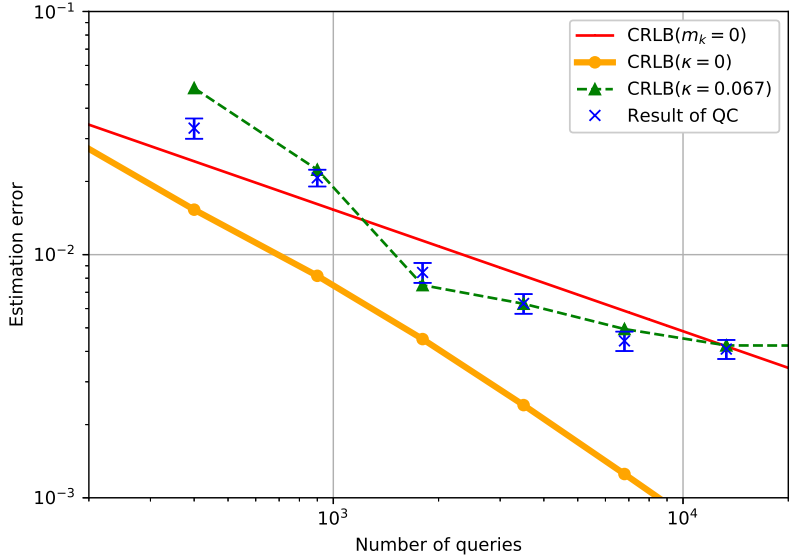}
	\caption{
		Estimation error of $a$ versus the total number of queries for the 2-qubit case.
		The thin red and thick yellow lines are the theoretical Cram\'er--Rao lower bound,
		obtained via the classical method and the ideal quantum ML method
		without noise ($\kappa=0$), respectively.
		The green dotted line shows the Cram\'er--Rao lower bound with noise $\kappa=0.067$.
		The blue cross marks show the standard deviation between the true value
		$S(f) = a = \sin^2\theta_a = 0.375$ and the estimated values of $a$ obtained via the
		ML method of experiments on \textit{ibmq\_valencia}.
	}
	\label{1248}
\end{figure}

We next experimentally executed the ML algorithm based on the model
\eqref{prob in Section 3.2}, for estimating $a=S(f)$ and $\kappa$.
Recall that the best ML estimate of $(a, \kappa)$ is given by the maximum of the likelihood
function \eqref{Likelihood in Section 3.2} with $P(m_k;a, \kappa)$ the model
\eqref{prob in Section 3.2} and $h_k$ the experimental result of the number of hit for a
fixed number of $\mathcal{Q}$ in amplitude amplification, $m_k~(k=0, \ldots, M)$;
here we tested 6 patterns $M = 1, \ldots, 6$.
In particular, we used EIS, fixed $N_k=100$ and $b=2\pi/5$ therefore $a=S(f)=0.375$.
The result is given in Fig.~\ref{1248}, which shows the relation between the estimation error
of $S(f)$ and the total number of query calls, $N_{\rm{q}}=\sum_{k=0}^{M}N_k(2 m_k+1)$.
The solid thin red and thick yellow lines are the theoretical Cram\'er--Rao lower bound
$\epsilon_{\rm min}(a,\kappa)$ given in Eq.~\eqref{2dim and 1dim CR inequality},
obtained via the classical method and the ideal quantum ML method
without noise ($\kappa=0$), respectively.
The blue cross marks are the standard deviation of the estimated values of $S(f)$ obtained
via the ML method, from the true value $S(f) = 0.375$.
Note that, for instance to mark the blue point at $M=4$ (or equivalently
$N_{\rm{q}} \sim 3.5\times 10^3$), in which case the estimation error is about
$0.65 \times 10^{-2}$, the ML algorithm uses the likelihood function constructed from the
amplitude amplification processes with different operation length $m_k~(k=0, \ldots, 4)$.
Further, to reduce the fluctuation of those points, we repeated the same experiment
$1,064$ times and averaged out to determine each point; 
the three-times standard errors are indicated by the error bars. 
The green dotted line shows the Cram\'er--Rao lower bound with noise level
$\kappa=0.067$, which is the single ML estimate of $\kappa$ based on the
$1,064 \times N_k =1,064 \times 100$ data at $M=6$ (that is, roughly speaking, the
\textit{best} estimate of $\kappa$ over the whole execution of the algorithm).

As seen from Fig.~\ref{1248}, the estimation error experimentally obtained using the ML
method (the blue points) is in good agreement with the theoretical Cram\'er--Rao lower
bound (the green dotted line).
A few slight deviation, particularly the points where the ML estimate is below the
Cram\'er--Rao lower bound, might be due to some imperfections other than depolarizing noise,
such as the rotation error of the gate operation and unnecessary coupling to neighboring
qubits on the device.
We would also like to emphasize that, in this example, there are several points where the
experiment achieves the more precise estimate than that obtained via the classical method
(the red line).
This is an important evidence that even a noisy quantum computer can be beneficial
over the classical one, in the measure of query complexity.
Finally we remark that a similar behavior was observed, in other settings that use different
two qubits in the device and a different target value $S(f)$;
see Appendix~\ref{sec:otherexperiment}.

\subsection{Experimental result for the three-qubit case}
\label{sec:3result}

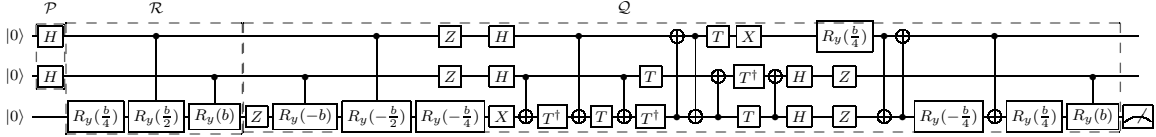
\begin{figure}[tb]
	\centering
	\resizebox{.99\linewidth}{!}{
	\Qcircuit @C=0.3em @R=.8em {
	&\mathcal{P} & & \mathcal{R} & &  & & & & & & & & & \mathcal{Q}\\
	\lstick{ \ket{0} } & \gate{H} & \ggate{}  & \ctrl{2} & \qw & \ggate{}  & \qw & \ctrl{2} & \gate{Z} & \gate{H} & \qw & \qw & \ctrl{2} & \qw & \qw & \qw & \targ & \ctrl{2} & \gate{T} & \gate{X} & \qw & \qw & \gate{R_y(\frac{b}{4})} & \ctrl{2} & \targ & \qw & \ctrl{2} & \qw & \qw & \qw \\
	\lstick{ \ket{0} } & \gate{H} &\qw & \qw & \ctrl{1} & \qw & \ctrl{1} & \qw & \gate{Z} & \gate{H} & \ctrl{1} & \qw & \qw & \qw & \ctrl{1} & \gate{T} & \qw & \qw & \targ & \gate{T^\dag} & \targ & \gate{H} & \gate{Z} & \qw & \qw & \qw & \qw & \qw & \ctrl{1} & \qw \\
	\lstick{ \ket{0} } & \qw & \gate{R_y(\frac{b}{4})} & \gate{R_y(\frac{b}{2})} & \gate{R_y(b)} & \gate{Z} & \gate{R_y(-b)} & \gate{R_y(-\frac{b}{2})} & \gate{R_y(-\frac{b}{4})} & \gate{X} & \targ & \gate{T^\dag} & \targ & \gate{T} & \targ & \gate{T^\dag} & \ctrl{-2} & \targ & \ctrl{-1} & \gate{T} & \ctrl{-1} & \gate{H} & \gate{Z} & \targ & \ctrl{-2} & \gate{R_y(-\frac{b}{4})} & \targ & \gate{R_y(\frac{b}{4})} & \gate{R_y(b)} & \meter
	\gategroup{2}{2}{3}{2} {.2em}{--}
	\gategroup{2}{3}{4}{5} {.2em}{--}
	\gategroup{2}{6}{4}{29} {.2em}{--}
	}
	}
	\caption{
		The 3-qubit circuit for computing the probability $P(m;a, \kappa)$ where the target value
		$S(f)$ is given in Eq.~\eqref{eq:SumSimpleSin} with $n=2$ and $b=2\pi/5$.
	}
	\label{3qubitsC}
\end{figure}

Here we present the experimental result for the case where the target integration
$\int_0^{1} \sin^2(bx) \ dx$ with $b=2\pi/5$ is to be approximated by $S(f)$
in Eq.~\eqref{eq:SumSimpleSin} with $2^n=2^2$ segments.
In this case, $S(f) = a = \sin^2\theta_a = 0.381$.
Also then we need $n+1=3$ qubits to implement the amplitude amplification operation.

The quantum circuit to execute the ML algorithm is shown in Fig.~\ref{3qubitsC}.
Because \textit{ibmq\_valencia} does not allow direct coupling for arbitrary pair of qubits, we chose three qubits to form a chain structure. In the experiment, the qubit in the middle of chain was chosen as the ancilla qubit, and it is
placed as the third qubit from the top in the circuit at Fig.~\ref{3qubitsC}.

\begin{figure}[tb]
	\centering
	\includegraphics[width=12.0cm]{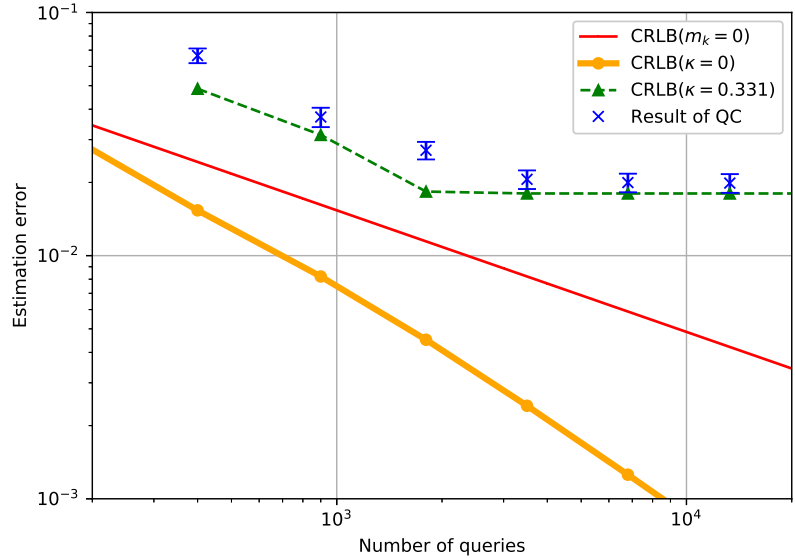}
	\caption{
		Estimation error of $a$ versus the total number of queries for the 3-qubit case.
		The thin red and thick yellow lines are the theoretical Cram\'er--Rao lower bound,
		obtained via the classical method and the ideal quantum ML method
		without noise ($\kappa=0$), respectively.
		The green dotted line shows the Cram\'er--Rao lower bound with noise  $\kappa=0.331$.
		The blue cross marks show the standard deviation between the true value
		$S(f) = a =\sin^2\theta_a = 0.381$ and the estimated values of $a$ obtained via the
		ML method of experiments on \textit{ibmq\_valencia}.
	}
	\label{3_1248}
\end{figure}

Fig.~\ref{3_1248} is the three-qubit version of that at Fig.~\ref{1248}, and shows
the relationship between the number of queries and the estimation error of $a=S(f)$.
The Cram\'er--Rao lower bound under noise (the green dotted line with triangle) was plotted
with $\kappa=0.331$.
Due to such a high noise level, as expected from Fig.~\ref{queryVSerror}, this
Cram\'er--Rao lower bound is above the classical one (the red line), meaning that the
quantum computation has no advantage in the amplitude estimation task.
However, what is important in our context is that the estimation error obtained from the
experiment (the blue points) lie near the green line; that is, the ML estimate computed
with the 3-qubit real quantum device asymptotically reaches the theoretical Cram\'er--Rao
lower bound.

Therefore, together with the result for the 2-qubit case, we now would like to conclude
that Eq.~\eqref{DC} is a good model of the noise process, at least for a small size qubit
device.
This means that the theoretical predictions illustrated in Fig.~\ref{queryVSerror} and thereby
Fig.~\ref{KVSerror} might be usable as a practical guide for discussing the condition on
the noise level $\kappa$ to realize the Heisenberg-scaling in the quantum amplitude
estimation task.

\section{Discussion}
\label{sec:discussion}

This section is divided to two topics.
In Section~\ref{sec:evaluation}, based on the result obtained in Section~\ref{sec:2.3}, we
show a procedure to assess the gate errors required to achieve a given task 
as well as the expected execution time of the algorithm,
and discuss the limitations such as the gate error and the coherence time when using IBM Quantum devices. 
Next in Section~\ref{sec:complexity_of_postprocessing}, we discuss the computational complexity that takes into account the classical optimization procedure to compute the ML estimate.

\subsection{Hardware specification for the amplitude estimation task}
\label{sec:evaluation}

Recall that Fig.~\ref{KVSerror} is used to predict the maximum allowed noise level
$\bar{\kappa}$ for achieving a given estimation error $\epsilon$ with the query calls
obeying the Heisenberg-scaling law.
Here we connect this value of $\bar\kappa$ to the error of elementary gates in
the quantum circuit; also the execution time of the algorithm is assessed.
The entire procedure to compute these quantities is composed
of the following four steps.
\begin{enumerate}

	\item
	      An amplitude estimation problem, together with a target estimation error $\epsilon$, is given to us.
	      Then the amplitude amplification operator $\mathcal{Q}$ (plus $\mathcal{P}$ and $\mathcal{R}$
	      in the Monte Carlo case), and accordingly the elementary gates constituting those operators
	      are identified; for instance, see Fig.~\ref{2qubitsC} or Fig.~\ref{3qubitsC}.

	\item
	      Given the target estimation error $\epsilon$, we use Fig.~\ref{KVSerror} to compute the
	      maximum allowed noise level $\bar{\kappa}$
	      when $N_k$ for all $k$ is fixed.
	      We can then compare $\bar\kappa$ to the ML estimate $\hat{\kappa}$ obtained through
	      the experiment with a real device, to see whether the device can produce the
	      desired estimate $\hat{a}$.
	      Also Eq.~\eqref{mlim} enables us to have the maximum number of operations of $\mathcal{Q}$,
	      i.e., $\bar{m}$, within which the Heisenberg-scaling nearly holds in estimating the parameter.
	      More precisely it is given by $\bar{m} = 0.5/(e^{\bar\kappa} -1)$;
	      for example, $\bar{m}$ can take only $\bar{m}=5$ when $\bar\kappa=0.1$, but $\bar{m}=500$
	      when $\bar\kappa=0.001$.

	\item
	      Furthermore, from $\bar{m}$ and the number of gates constituting the operator
	      $\mathcal{Q}$, denoted by $L$, we can have a rough estimate of the execution time of each
	      quantum circuit with length $m_1, m_2, \ldots, \bar{m}$.
	      Then the execution time in $\bar{m}$ and
	      the total execution time in $m_1, m_2, \ldots, \bar{m}$
	      can also be estimated, which are then compared to the coherence
	      time of the available real device, and then the feasibility of this algorithm can be assessed respectively.

	\item
	      Under the assumption that all qubits are subjected to only depolarizing noise, we approximate
	      the $p=e^{-\bar\kappa}$, the success probability of the operator $\mathcal{Q}$, as
	      $\prod_{i=1}^{L} (1 - \epsilon_i)$, where $L$ is
	      the number of gates that constitute the operator $\mathcal{Q}$ and $\epsilon_i$ is the
	      depolarizing error probability of the $i$th qubit.
	      Using this equation we may also have a rough estimate of the gate error, using $\bar{\kappa}$
	      and $L$.
	      The gate error is compared to that of real device (which might be identified via the standard
	      randomized benchmarking test) to see the feasibility of the algorithm.

\end{enumerate}
A detailed procedure for computing the above quantities, especially the gate errors and
the total execution time, is given in Appendix~\ref{sec:cal}.

As an example, let us consider a multiple integration over 5 variables, which assumes direct
correlations between any of two variables, e.g., $\int (x_1x_2 + x_2x_3 + x_3x_4 + x_4x_5)d\bf{x}$.
The reason of this choice is that, to compute such a multiple integration with more than five
variables, the Monte-Carlo method is usually employed rather than grid-based numerical
integration approaches.
Here we follow the above four steps one by one.
(Step 1)
Suppose that we are required to achieve the target estimation error $\epsilon = 0.001$ with the
Heisenberg-scaling query calls.
(Step 2)
Then Fig.~\ref{KVSerror} tells us that we need $\bar{\kappa} = 0.005$ when $N_k=100$ for all $k$ and 99 qubits wherein
$48$ qubits are used for the ancilla to gate decomposition.
Also we have $\bar{m}=99$.
(Step 3)
For some quantum devices, the operating time of single gates and the CNOT gate are
publicly available; in the case of {\it ibmq\_valencia}, they are $7.1 \times 10^{-8}$ s and
$2.8 \times 10^{-7}$ s, respectively.
Also the measurement time is $3.5 \times 10^{-6}$ s.
Moreover we assume that the interval time between the measurement and the next execution
of the algorithm is 10 times longer than the execution time of the algorithm with length $\bar{m}$.
Then, we find that the execution time of the algorithm with length $\bar{m}$ is $0.54$ s
and the total computation time of the entire algorithm is about $1,082$ s;
see Table~\ref{tab:assumptions} in Appendix~\ref{sec:cal} for the detailed calculation.
(Step 4)
We can also assess the gate error required for the algorithm to follow the Heisenberg-scaling
to reach the given estimation error.
Under the assumption that the error of CNOT gate, $\epsilon_{\rm d}$, is 10 times bigger than
that of any single gate error, $\epsilon_{\rm s}$, we end up with
$\epsilon_{\rm d}=2.8 \times 10^{-7}$ and $\epsilon_{\rm s}=2.8 \times 10^{-8}$.
Now, in the case of {\it ibmq\_valencia}, they are given by $\epsilon_{\rm d}\sim  1.0 \times 10^{-2}$
and $\epsilon_{\rm s}\sim 1.0\times 10^{-3}$.

An important finding is in quantifying the difference between the gate error of the currently available devices and that of the near ideal machine which realizes the Heisenberg-scaling to reach the given estimation error. The execution time (0.54 s) is also clarified, which may be much longer than the coherence time of the currently available devices.
In fact, even though these gaps seem to be large, they are informative for us because we now know how much improvement in the hardware development is necessary to fill these gaps.
In addition, algorithm improvements would help close these gaps.
For example, recent research~\cite{vazquez2020efficient} suggested a method for reducing the circuit depth and the number of qubits.

Finally, we note that the estimated noise level $\kappa$ is comparable with 
the value calculated based on the publicly available information on the gate 
error of the IBM Quantum device. 
In our case, the 2-qubits Grover circuit contains 5 CNOT gates (the $Q$ operator 
part of Fig. \ref{2qubitsC}), and the 3-qubits one contains 16 CNOT gates (the 
$Q$ operator part of Fig. \ref{3qubitsC}); note that $R_y$ gate is composed of 
2 CNOT gates. 
Then, for the 2 qubits case, the CNOT gate error was 0.00565, meaning that 
$\kappa \sim -{\rm ln}((1-0.00565)^5) \sim 0.0283$. 
For the 3 qubits case, we used two types of CNOT gate with error rate 
0.008923 and 0.01119, and they were used 8 times; hence we have 
$\kappa \sim -{\rm ln}((1-0.008923)^8*(1-0.01119)^8) \sim 0.162$. 
By taking into account some other impact of noise such as single gate error, 
we consider that these values becomes closer to the estimated values 
$\kappa = 0.067$ and $\kappa = 0.331$.

\subsection{Computational time complexity of maximum likelihood estimation\label{sec:complexity_of_postprocessing}}

We have argued about the possibility of less number of queries achieved by the ML method against classical cases under noise influences,
and how to deal with anomalous target values. Here, assuming anomalous targets are detectable, we show the classical post-processing for maximum likelihood
estimation of the target value $\hat{a}$ and noise level $\hat{\kappa}$ can be performed in $O(\ln^{5/2}(1/\epsilon))$, which is much less than that of the quantum part.

Because under depolarizing noise model the success probability now has two parameters, the target value $\hat{a}$
and noise level $\hat{\kappa}$ as shown in Section~\ref{Section 2.2}, it is necessary to perform
two-dimensional maximum likelihood to obtain the estimated values of the parameters.
A straightforward random search or grid search requires $O(1/\epsilon^2)$ evaluations
of likelihood function to achieve the error $\epsilon$~\cite{zabinsky2013stochastic}. This completely
ruins the advantage obtained from the quantum method because the total time complexity now becomes
$O(1/\epsilon^2)$ which is at least the same as the classical Monte-Carlo approach.

In order to avoid the problem of computational complexity, we employed an adaptive constant grid search
at each stage $k$ for $0 \le k \le M$ in the experiment of Section~\ref{sec:result}. Namely, at the $k$-th stage,
we performed the grid search with a constant number of divisions only in the range of confidence interval defined as $C_\epsilon$ times larger than the error estimated from the Fisher information at the $(k-1)$-th stage. By the Chebysev inequality, the error estimated at each stage is guaranteed to be within $C_\epsilon$ times the confidence interval with probability at least $1 - 1/C_\epsilon^2$. To obtain an estimation within $\epsilon$ error, the number of stages $M$ is $O(\ln(1/\epsilon))$.
Thus, the total probability of obtaining good parameter estimation is at least $(1-1/C_\epsilon^2)^M$, which is $\Omega(1)$ when $C_\epsilon = \Theta(\ln^{1/2}(1/\epsilon))$. The total time complexity of the
maximum likelihood computation is thus $C_\epsilon \cdot M \cdot O(\ln(1/\epsilon))$, where the last term is that for evaluating the likelihood
function. This gives the bound $O(\ln^{5/2}(1/\epsilon))$ time complexity for the classical post processing.

Except for the anomalous cases, we should note the estimation error of the target value $a$ is not affected even if the estimation error of the noise level $\kappa$ is large. The details are shown in Appendix~\ref{sec:effectOfKappaToTheta}.
This may allow us to reduce the two-dimensional ML estimation to a one-dimensional one. In this case, the estimated values can be obtained with lower computational cost; roughly estimate the noise level $\kappa$ with large grid size, and then perform one-dimensional ML of the target value $a$ with high accuracy. Furthermore, for sufficiently large number of shots the two-dimensional ML estimation
can be approximated with the weighted least square estimation. The latter can be
solved much more efficiently in practice thanks to the specific algorithms, such as the Levenberg-Marquardt's \cite{nocedal2006numerical,marquardt1963algorithm}.

\section{\label{sec:con} Conclusion}

All quantum algorithms running on currently available quantum devices must consider the effects of noise.
Hence, to perform an appropriate evaluation of such algorithms, particularly those aimed at possible quantum advantage, it is necessary to carefully model the noise, analyze its effects to the algorithm, and make a validation via experiments.
This paper provides a demonstration of this evaluation, yielding Fig. \ref{KVSerror},
a relationship between the target estimation error and the required noise threshold for the amplitude estimation problem, which is validated by experiments on quantum devices.

Lastly we again point out the importance of multi-parameters estimation problem 
in the quantum computing framework. 
In our case, the nuisance noise parameter in addition to the target amplitude parameter 
must be estimated. 
We have seen that the Fisher information matrix can be degenerate, in which case the target amplitude parameter cannot be efficiently estimated even by increasing the 
number of amplitude amplifications. 
This singular problem arises only in multi-parameters estimation problems, and 
a similar challenge may easily appear in other scenarios such as the phase estimation problem under noisy environment. 
In this sense, while this paper employed the ML estimate approach, we can have 
various options to design an efficient estimator for such multi-parameter estimation 
problems in noisy quantum devices. 
For instance Ref.~\cite{wang2020bayesian} showed the method to tune the likelihood 
function for Bayesian inference of the amplitude parameter, and they have derived 
the run time estimation to achieve a target accuracy under noisy environment.

\begin{acknowledgements}

This work was supported by MEXT Quantum Leap Flagship Program Grant Number JPMXS0118067285 and JPMXS0120319794.
The authors acknowledge helpful discussions with Takahiko Satoh for designing 
the quantum circuit, and thank Naoki Kanazawa for useful comments about IBM Quantum Systems.
The results presented in this paper were obtained in part using an IBM Quantum quantum 
computing system as part of the IBM Quantum Network. The views expressed are those of 
the authors and do not reflect the official policy or position of IBM or the 
IBM Quantum team.
\end{acknowledgements}

\begin{appendices}

\section{Effect of estimation error of noise level on the target value\label{sec:effectOfKappaToTheta}}

In this paper the problem is formulated as a two-parameters
estimation problem with respect to $(a, \kappa)$, where $a$
is our main interest while $\kappa$ is the nuisance parameter.
Hence here we briefly sketch how much the estimation error of
$\kappa$ may affect on the target value $a$.

For this purpose, let us consider one-dimensional ML estimation
for the target $a$, with a fixed noise level $\hat{\kappa}$, which
is supposed to be roughly estimated in advance in some way.
The true values of parameters, $a$ and $\kappa$, are denoted as
$a_0$ and $\kappa_0$ only in this section, and the differences
between the true values and the estimation errors are denoted as $\delta a = \hat{a}-a_0$ and $\delta \kappa = \hat{\kappa}-\kappa_0$, respectively.
By definition, the one-dimensional ML estimate $\hat{a}$ is given
by
\begin{equation}
	\hat{a}
	=\argmax_{a}\ln L(\mathbf{h};\{a,\hat{\kappa}\}),
	\label{eq:maximize_theta_1dim}
\end{equation}
where $L$ is the likelihood function introduced in Eq.~\eqref{likelihood}.
Assuming the smoothness of the likelihood function with respect to $a$, we have the following equation:
\begin{equation}
	\frac{\partial}{\partial a}\ln L(\mathbf{h};\{\hat{a},\hat{\kappa}\})=0.
\end{equation}
Hence, the first order Taylor expansion of the left-hand side of
this equation, at around the true parameters $a_0$ and $\kappa_0$,
yields
\begin{equation}
	\begin{split}
		\frac{\partial}{\partial a}\ln L(\mathbf{h};\{a_0,\kappa_0\}) =   -\delta a \frac{\partial^2}{\partial a^2}\ln L(\mathbf{h};\{a_0,\kappa_0\}) -\delta \kappa  \frac{\partial^2}{\partial a \partial \kappa}\ln L(\mathbf{h};\{a_0,\kappa_0\}).
	\end{split}
	\label{eq:expansion_ML}
\end{equation}
As a consequence of the central limit theorem, the left-hand side of this equation asymptotically follows $\mathbb{N}(0,\mathcal{I}_{1,1})$,
i.e., the normal distribution with mean $0$ and variance $\mathcal{I}_{1,1}$, in the limit with a large number of shots.
Also, the first term on the right-hand side asymptotically converges to $ \mathcal{I}_{1,1}\delta a $ and the second term converges to $ \mathcal{I}_{1,2}\delta \kappa $.
Thus in the asymptotic regime, Eq.~\eqref{eq:expansion_ML} implies
that $\delta a$ obeys the normal distribution with mean
$-\mathcal{I}_{1,2} \delta \kappa / \mathcal{I}_{1,1}$ and
variance $1/\mathcal{I}_{1,1}$, i.e.,
\begin{equation}
	\delta a \sim \mathbb{N} \left( -\frac{\mathcal{I}_{1,2}}{\mathcal{I}_{1,1}} \delta \kappa, \frac{1}{\mathcal{I}_{1,1}}\right).
\end{equation}
Then the mean squared error of $\hat{a}$ can be calculated as
\begin{equation}
	\mathbb{E} \left[ \delta a^2 \right] = \delta \kappa^2\frac{\mathcal{I}_{1,2}^2}{\mathcal{I}_{1,1}^2}  +  \frac{1}{\mathcal{I}_{1,1}}.
\end{equation}
If the squared error of $\hat{\kappa}$ is $c$ times bigger than
the Cram\'{e}r--Rao lower bound of the variance,  $\mathcal{I}_{1,1}/(\mathcal{I}_{1,1}\mathcal{I}_{2,2} -\mathcal{I}_{1,2}^2)$, this equation yields
\begin{equation}
	\begin{split}
		\mathbb{E} \left[ \delta a^2 \right] =  c \frac{\mathcal{I}_{1,1}}{(\mathcal{I}_{1,1}\mathcal{I}_{2,2} -\mathcal{I}_{1,2}^2)}\frac{\mathcal{I}_{1,2}^2}{\mathcal{I}_{1,1}^2}  +  \frac{1}{\mathcal{I}_{1,1}}
		= (\mathcal{I}^{-1})_{1,1}
		\left(1+(c-1)\frac{\mathcal{I}_{1,2}^2}{\mathcal{I}_{1,1}\mathcal{I}_{2,2}}\right).
	\end{split}
\end{equation}
The first term of the rightmost side of this equation is just the Cram\'{e}r--Rao lower bound and the second term is an additional term due to the effect from the error of noise parameter $\kappa$.
This expression indicates that the estimation error of $\kappa$
does not significantly affect on the estimation error of the
target value $a$ except for the anomalous target values, because
$\mathcal{I}_{1,2}^2/\mathcal{I}_{1,1}\mathcal{I}_{2,2}$ is not
large for such typical target values.

\section{Detailed procedure for determining the hardware components}
\label{sec:cal}

In Section~\ref{sec:evaluation} we have discussed the method to evaluate the hardware components required to achieve the Heisenberg-scaling estimation under the influence of noise.
Here we show a detailed procedure to determine those quantities
for an $N_{\rm int}$-dimensional integration problem, which is
summarized in Table~\ref{tab:assumptions}.
Our task is to use the quantum ML method to estimate the value
of integral, within the target estimation error $\epsilon$.
Note that then the amplitude amplification (AA) operator
$\mathcal{Q}$ as well as $\mathcal{R}$ and $\mathcal{P}$ are determined.
Also we set the following assumptions;
the gate error of CNOT, $\epsilon_d$, is 10 times bigger than that
of any single gate error, $\epsilon_s$;
the qubits are fully connected in the device;
the operating time of single and CNOT gates are $7.1\times 10^{-8}$ s and  $2.8\times 10^{-7}$ s, respectively;
the measurement duration time is $3.5\times 10^{-6}$ s;
the interval time between the measurements is $10$ times bigger
than the execution time;
the number measurement of $N_k=100$ and the sequence is EIS.
The main focused quantities are the gate errors
$(\epsilon_d, \epsilon_s)$ and the executing time of the algorithm.

\begin{table}[htb]
	\caption{List of quantities to be determined}
	\label{tab:assumptions}
	\begin{center}
		\resizebox{.99\linewidth}{!}{
			\begin{tabular}{lccl}
				\hline
				\multicolumn{1}{c}{element}                                               & calculated by                                                              & value               & \multicolumn{1}{c}{supplement} \rule[-8pt]{0pt}{20pt} \\
				\hline
				$N_{\rm nq}$ (\# qubits for an integral)                                  & $\log_2 {1/\epsilon}$                                                      & $10$                & $\epsilon=0.001$ \rule[-8pt]{0pt}{20pt}               \\
				$N_{\rm tnq}$ (total number of qubits)                                    & $2 N_{\rm nq} N_{\rm int} -1$                                              & $99$                &
				\begin{tabular}{l}
					$N_{\rm nq} N_{\rm int} +1$ uses ML method                \\
					$N_{\rm nq} N_{\rm int} -2$ comes from the ancilla qubits \\
					for decomposition of $\mathbf{S}_0$
				\end{tabular}                                                                                                                                                                                                           \\
				$\bar{\kappa}$ (noise level)                                              & $5 \epsilon /\sqrt{N_k} $                                                  & $0.005$             & \ From Fig.~\ref{KVSerror}. \rule[-8pt]{0pt}{20pt}    \\
				$N_{y}$ (number of $CCR_y$)                                               & $ N_{\rm nq}(N_{\rm nq}-1)N_{\rm int}^2/2$                                 & $1,000$             & \  From $\mathcal{R}$ \rule[-8pt]{0pt}{20pt}          \\
				$N_s$ (\# single gates in $\mathcal{Q}$)                                  & $2(N_{\rm nq}N_{\rm int}+1+6N_{y})+12N_{\rm nq}N_{int}-15$                 & $12,687$            &
				\begin{tabular}{l}
					$2(N_{\rm nq}N_{\rm int}+1)$ is from $\mathcal{P}$ and $\mathcal{P}^{-1}$ \\
					$2(6N_{y}*2)$ is from $\mathcal{R}$ and $\mathcal{R}^{-1}$                \\
					$12 N_{\rm nq} N_{\rm int} - 15$ is from decomposition of $\mathbf{S}_0$
				\end{tabular}                                                                                                                                                                                                           \\
				$N_d$ (\# CNOT gates in $\mathcal{Q}$)                                    & $8N_{y}*2+6N_{\rm nq}N_{\rm int}-5$                                        & $16,295$            &
				\begin{tabular}{l}
					$8 N_{y}\times 2$ is from $\mathcal{R}$ and $\mathcal{R}^{-1}$ \\
					$6 N_{\rm nq} N_{\rm int} - 5$ is from decomposition of $\mathbf{S}_0$
				\end{tabular}                                                                                                                                                                                                           \\
				\begin{tabular}{l}
					$\epsilon_s$ (single gate error) \rule[-6pt]{0pt}{15pt} \\ $\epsilon_d$ (CNOT gate error) \rule[-6pt]{0pt}{15pt}
				\end{tabular}                                                &
				$\mathrm{e}^{-\bar{\kappa}} = (1-\epsilon_s)^{N_s} (1-\epsilon_d)^{N_d} $ &
				\begin{tabular}{c}
					$2.8\times 10^{-8}$ \rule[-6pt]{0pt}{15pt} \\ $2.8\times 10^{-7}$ \rule[-6pt]{0pt}{15pt}
				\end{tabular}                                                &
				\begin{tabular}{l}
					Assuming $\epsilon_s = 0.1\epsilon_d$ and \\
					including approximation calculations.
				\end{tabular} \rule[-12pt]{0pt}{30pt}                                                                                                                                                                                   \\
				$\bar{m}$ (maximum number of $\mathcal{Q}$)                               & $(2 \bar{m}+1) (1-\mathrm{e}^{-\bar{\kappa}})<1 $                          & $99$                & \ From Eq.~\eqref{mlim} \rule[-8pt]{0pt}{20pt}        \\
				$t_{AA}$ (the time of a $\mathcal{Q}$)                                    & $t_{AA}= t_{s} N_s + t_{d} N_d$                                            & $5.4\times 10^{-3}$ &
				\begin{tabular}{l}
					$t_s$, $t_d$ are the gate time of single and                \\
					CNOT gate respectively which are                            \\
					$7.1\times10^{-8}$, $2.8\times 10^{-7}$ seconds from IBM Q. \\
				\end{tabular}                                                                                                                                                                                                           \\
				$t_{\bar{m}}$ (executing time with $\bar{m}$)                             & $t_{AA} \times \bar{m}+ t_m$                                               & $0.54$              &
				\begin{tabular}{l}
					$t_m$ is the measurement time, $3.5 \times 10^{-6}$ seconds.
				\end{tabular} \rule[-8pt]{0pt}{20pt}                                                                                                                                                                                    \\
				$t_t$ (total executing time)                                              & $\displaystyle \sum_{k=1}^{N}\left( (t_{AA} m_k+ t_m +t_i) * N_k \right) $ & $1,082$             &
				\begin{tabular}{l}
					$N$ is from $m_N = \bar{m}$.                       \\
					$t_i$ is an interval time between shots calculated \\
					as $10$ times the execution time.                  \\
					$N_k=100$ is the number of shots, for all $k$      \\
				\end{tabular}                                                                                                                                                                                                           \\
				\hline
			\end{tabular}
		}
	\end{center}
\end{table}

\begin{figure}[!tb]
	\begin{minipage}{\hsize}
		\begin{center}
			\includegraphics[width=10.0cm]{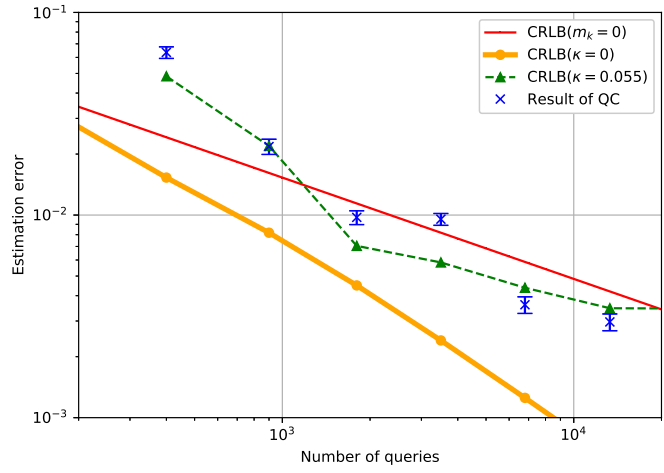}
		\end{center}
		\subcaption{The case of target value $S(f) = a = 0.375$}
		\label{fig:one}
	\end{minipage}
	\begin{minipage}{\hsize}
		\begin{center}
			\includegraphics[width=10.0cm]{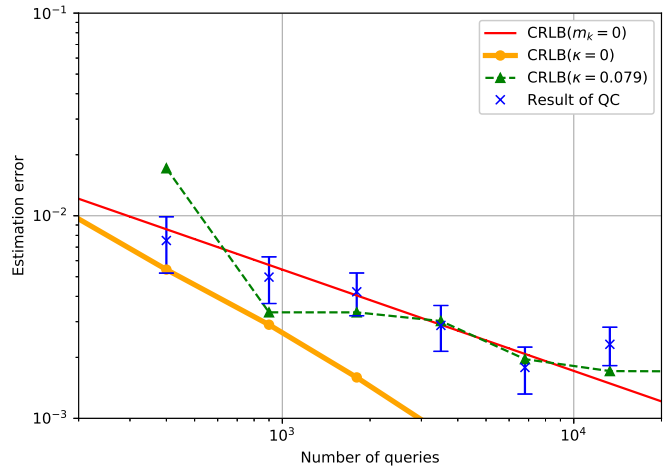}
		\end{center}
		\subcaption{The case of target value $S(f) = a \sim 0.03$}
		\label{fig:two}
	\end{minipage}
	\caption{Estimation error of $a$ versus the total number of queries.
		These figures have different target values, while the other settings are the same.
		The thin red, thick yellow, and green dotted lines are the theoretical Cram\'er--Rao lower bounds,
		while the blue error bars represent the three-times standard errors.
		For the detail, see Section~\ref{sec:2result}.}
\end{figure}

\section{Two qubit experimental result with other settings\label{sec:otherexperiment}}

In Section~\ref{sec:2result}, we demonstrated the results using the real quantum computer with specific parameters.
Here we present additional results in different experimental settings, to see how the basic quantities such as the estimation
error would be affected by those changes.
Fig.~\ref{fig:one} shows the result when we used the third and the fourth qubits on \textit{ibmq\_valencia} to represent the target value $S(f) = a = 0.375$, whereas the other settings are the same as before; e.g., EIS ($m_0=0, m_k=2^{k-1}$) is taken and the number of shots is $N_k=100\ \forall k$.
Next, Fig.~\ref{fig:two} shows the case where the target value is chosen as $S(f) = a=\sin^2 \pi/40+\sin^2 3\pi/40 \sim 0.03$, meaning that $b=\pi /10$; the other settings are the same as in
Fig.~\ref{fig:one}.

These figures show that
the estimation error obtained with the experiments (the blue cross marks) lies near 
the green dotted line which is the Cram\'{e}r--Rao lower bound with depolarizing noise, supporting our depolarizing noise model discussed in Section~\ref{sec:2.3}. 
We note that the noise level $\kappa$ ended up with different values, $0.055$ 
and $0.079$ respectively, depending on the different target values $a$; 
hence it is important to perform the multi-parameter estimation each time even when 
using the same qubits. 
In addition, the experimental results of Fig.~\ref{1248} and Fig.~\ref{fig:one} show 
that the noise level $\kappa$ differs depending on the qubits; hence, a careful qubit 
mapping (i.e., assigning the location of qubits) is important to get to lower noise 
level.

Note also that, in Figs. 11(a) and (b), the error bars (the three-times standard 
errors) seem to have different values, but this is merely because the vertical axis 
is in the log scale and each difference is only within a factor of two. 

\end{appendices}

\bibliographystyle{spmpsci_unsort}      
\bibliography{Manuscript} 

\end{document}